\documentclass[12pt]{article}
\usepackage{amsmath,amssymb,amsfonts}
\usepackage[dvips]{graphicx}
\usepackage{epsfig}

\makeatletter \@addtoreset{equation}{section} \makeatother
\makeatletter \@addtoreset{figure}{section} \makeatother

\def\CE{{\cal E}}\def\CF{{\cal F}}
\def\CH{{\cal H}}\def\CI{{\cal I}}
\def\CK{{\cal K}}\def\CL{{\cal L}}\def\CM{{\cal M}}
\def\CN{{\cal N}}\def\CO{{\cal O}}
\def\CQ{{\cal Q}}
\def\CU{{\cal U}}\def\CV{{\cal V}}
\def\CW{{\cal W}}\def\CX{{\cal X}}

\def\a{\alpha}\def\b{\beta}\def\g{\gamma}
\def\d{\delta}\def\e{\epsilon}
\def\z{\zeta}\def\th{\theta}
\def\l{\lambda}

\def\s{\sigma}
\def\t{\tau}
\def\G{\Gamma}

\def\O{\Omega}
\def\vare{\varepsilon}

\def\da{{\dot{\a}}}

\begin{document}
\begin{titlepage}
\vfill
\begin{flushright}
{\tt\normalsize DAMTP-2010-123}\\
{\tt\normalsize KIAS-P11003}\\
\end{flushright}
\vfill
\begin{center}
{\large\bf Framed BPS States, Moduli Dynamics, and Wall-Crossing
}

\vfill

Sungjay Lee\footnote{\tt s.lee@damtp.cam.ac.uk}$^\diamondsuit$
and Piljin Yi\footnote{\tt piljin@kias.re.kr}$^\spadesuit$

\vskip 5mm
$^\diamondsuit${\it DAMTP, Centre for Mathematical Sciences, University of Cambridge, }\\
{\it Wilberforce Road, Cambridge, CB3 0WA, UK} \\
\vskip 5mm
$^\spadesuit${\it  School of Physics, Korea Institute for Advanced Study, Seoul 130-722, Korea}

\end{center}
\vfill

\begin{abstract}
\noindent
We formulate supersymmetric low energy dynamics for  BPS dyons in strongly-coupled  $N=2$
Seiberg-Witten theories, and derive wall-crossing formulae thereof. For
BPS states made up of a heavy core state and $n$ probe (halo) dyons around it, we
derive a reliable supersymmetric moduli dynamics with $3n$ bosonic coordinates and
$4n$ fermionic superpartners. Attractive interactions are captured via
a set of supersymmetric potential terms, whose detail depends only on the
charges and
the special K\"{a}hler data of the underlying $N=2$ theories. The
small parameters that control the approximation are not electric couplings
but the mass ratio between the core and the probe, as well as
the distance to the marginal stability wall where the central charges
of the probe and of the core align. Quantizing the dynamics, we construct
BPS bound states and  derive the primitive and the semi-primitive
wall-crossing formulae from the first principle. We speculate on
applications to line operators and Darboux coordinates, and also about
extension to supergravity setting.
\end{abstract}

\vfill
\end{titlepage}

\parskip 0.1 cm
\tableofcontents\newpage
\renewcommand{\thefootnote}{\#\arabic{footnote}}
\setcounter{footnote}{0}

\parskip 0.2 cm
\section{Introduction}

The wall-crossing in supersymmetric theories refers to the
phenomenon where certain one-particle BPS states \cite{Prasad:1975kr,Bogomolny:1975de}
disappear from the
spectrum as the vacuum moduli or parameters are changed
continuously. The naive stability argument of BPS states
relies on the short-multiplet structure due to partially
preserved supersymmetry, but this is really applicable only
when we consider dynamical processes in a given vacuum. When we
change vacuum or parameters, even continuously, the state
itself can disappear from the spectrum altogether at which
point the supermultiplet structure of the state becomes
a moot issue.

Although the wall-crossing had an early precursor
in the context of supersymmetric kinks in two-dimensional
${N} = (2,2)$ theories \cite{Cecotti:1992rm,Cecotti:1992qh},
it was in the context of ${N} = 2$
supersymmetric theories in four dimensions, such as
Seiberg-Witten theory \cite{Seiberg:1994rs,Seiberg:1994aj}
and Calabi-Yau compactification
of type II string theories, that the phenomenon came under wide
scrutiny. The co-dimension one surface across which a BPS state
disappear is called the marginal stability wall (MSW),
and their presence renders the problem of finding BPS spectrum
extremely complicated.
For the simplest of Seiberg-Witten theories, monodromy
properties \cite{Seiberg:1994rs} alone can determine the spectrum \cite{Ferrari:1996sv}
but this is more an exception than a rule.

Despite such early discoveries, the space-time picture of exactly what
happens to the state upon the wall-crossing remained unclear until
it was uncovered in the 
context of
1/4 BPS dyons in $N=4$ super Yang-Mills theory, which  preserve
four supersymmetries just as  1/2 BPS objects of $N=2$ theories do.
It was found in Ref.~\cite{Lee:1998nv} that such BPS dyons
must be, typically, thought of as a loose bound states of more than one
dyonic centers with mutually non-local charges. The
distances between such centers are not free but determined
by the vacuum moduli $u_i$'s
\begin{equation}\label{old}
R_{AB}=R_{AB}(u_i;\{q^i_C,p^i_C\}) \ ,
\end{equation}
with $(q^i_A,p^i_A)$ being the charges of the $A$-th center.
In particular, some of $R_{AB}$ was shown to diverge as a MSW is
approached; this happens simply because
scalar forces and electromagnetic forces do not cancel each other between
dyons of mutually nonlocal charges and the equilibrium distance
is determined by a  detailed balance of classical forces: state by state,
the wall-crossing has a very mundane and classical explanation.

This finding is immediately applicable to weakly coupled $N=2$ theories
as well, because $N=2$ theory BPS solitons can  be classically embedded
to a $N=4$ theory. There will be differences at quantum level because
the supermultiplet structures  (and flavor structures) are different,
but the above space-time picture of wall-crossing is essentially classical
and quite robust.

For both $N=2$ and $N=4$ theories, this multi-center nature and
the subsequent wall-crossing were soon elevated to the semiclassical
level \cite{Bak:1999da,Gauntlett:1999vc,Bak:1999vd}.
The quantum low energy dynamics of magnetic monopoles was derived
rigorously from the super Yang-Mills theories in question \cite{Gauntlett:2000ks},
and dyons were constructed as quantum bound states of monopoles with
certain conjugate momenta turned on \cite{Bak:1999ip,Gauntlett:1999vc}.
What used to be the classical orbit size is now represented by
the quantum bound state size, and is still determined by
the vacuum moduli and  charges as in (\ref{old}). The size of
the bound state is again divergent as a wall of marginal stability
is approached, across which the state no longer exists as quantum
and BPS  one-particle state.

In such supersymmetric low energy dynamics of solitons, precise state
counting is a simple matter of finding bound state wavefunctions or
computing index of ceratin Dirac operators on the moduli space.
For instance, the bound state of a pair of dyons of charge
$\g_1+\g_2$ has been constructed and counted when the total magnetic charge
is a dual root. The degeneracy on one side of
a wall \cite{Gauntlett:1999vc}\footnote{In this note, we
take the convention that  Schwinger
products take values in ${\mathbf Z}/2$. }
can be written as
\begin{equation}
|\,\O(\g_1+\g_2)|=2|\langle \g_1,\g_2\rangle| \ ,
\end{equation}
where we introduced  $\Omega(\gamma)$, the second
helicity trace for the supermultiplet of charge $\gamma$, as
\begin{align}\label{2ndh}
  \O = -\frac12 \text{tr}\big( (2J_3)^2 (-1)^{2J_3} \big)\ .
\end{align}
A simplest generalization of this is a chain of dyons with
nearest neighbor interactions,
namely $\langle \g_A,\g_B\rangle\neq 0$ if and only if $|A-B|=1$.
Whenever such a state exists as a quantum BPS state, the
degeneracy takes a simple form \cite{Stern:2000ie},
\begin{equation}
|\,\O(\g_1+\g_2+\g_3+\cdots)|=\Big|\prod_A 2\langle \g_A,\g_{A+1}\rangle\Big| \ .
\end{equation}
They were shown to exist only when each and every one of
$\langle \g_A,\g_{A+1}\rangle$ obeys certain inequalities
defined by the vacuum moduli, which amounts to being on the ``right"
side of several MSW's, basically one for each interacting pair.
The formula clearly suggests that such states
can be constructed iteratively by attaching one kind of dyons at a time,
already hinting at a simple universal wall-crossing formula.

The next breakthrough came from ${N}=2$ supergravity
analysis by Denef who also found the multi-centered nature of BPS
black holes and the subsequent wall-crossing in the context of
attractor flow solutions \cite{Denef:2000nb,Denef:2001xn}.
The approach gave a universal and
explicit constraints for the relative positions of charge
centers, say,  charge $\gamma_A$  at $\vec x_A$,
\begin{eqnarray}\label{denef}
\sum_{B\neq A}\frac{\langle \gamma_B,\gamma_A\rangle}{|\,\vec x_B-\vec x_A|}
={\text{Im}\big[ \zeta_T^{-1} Z(\g_A) \big]}\ .
\end{eqnarray}
where $Z(\g_A)$ is the central charge of $\gamma_A$
and $\zeta_T $ is the phase factor of the total central charge
$Z_T=\sum_A Z(\g_A)$. In supergravity, this simplifies and
supersedes the field theory results which we abstractly noted
as (\ref{old}).

The wall-crossing for supergravity black hole solutions is again
due to a divergent distance between the charge centers, which is
dictated by long distance classical physics, just as
as in the field theory soliton picture of BPS dyons:
the sign of the left hand side  of (\ref{denef}) is independent of vacuum moduli
while that of the right hand side can flip the sign as we change
vacuum. Clearly at some point where the right hand side approaches
zero from the positive side, one of the distances has to diverge,
beyond which the solution can no longer exist. This is most useful since
the sizes of the states can be found without detailed construction.
However,  there is no information on
how a given charge state is split into what charge centers,  unlike the
explicit constructions of multi-center solution/quantum states
in the field theory story.

Although supergravity solutions themselves were not amenable to
explicit and precise quantum counting,
Denef further went on to conjecture general two-body
wall-crossing formula that extends the above field theory
result to arbitrary (magnetic) charges \cite{Denef:2002ru}. With spin content
taken into account, the formula reads,
\begin{equation}
\O(\g_1+\g_2)=-(-1)^{2|\langle \g_1,\g_2\rangle|} \,2|\langle \g_1,\g_2\rangle|\,\O(\g_1)\,\O(\g_2)\ ,
\end{equation}
which was later extended by Denef and Moore to the
semi-primitive cases \cite{Denef:2007vg}, captured in a generating function,
\begin{eqnarray}\label{gen}
   \sum_{n=0} \O(\g_1+n \g_2) \,q^n = \O(\g_1)\prod_{k=1} \
  \Big[ 1 - (-1)^{2k\langle \g_2, \g_1 \rangle} q^k \Big]^{
  2 k | \langle \g_2, \g_1 \rangle| \O(k\g_2) }  \ ,
\end{eqnarray}
counting the BPS states of charges $\gamma_1+n\gamma_2$
in terms of degeneracies of states with charges $\gamma_1$ and $n\gamma_2$.
These spurred much activities toward general solutions to the wall-crossing
problem, and was integrated recently into
more general Kontsevich-Soibelman's wall-crossing formalism \cite{KS}.

Despite evidences that support the semi-primitive
wall-crossing formulae of Denef-Moore (which in turn support
Kontsevich-Soibelman formalism), it has been rigorously tested only in
 specific cases. The most systematic example of
this is the ${N}=2$ weak coupling analysis that preceded the conjecture,
but the limitation of weak coupling limit casts some shadows
on its general usefulness. It would be very useful if we can
find a similarly systematic method of constructing and counting
BPS bound states, and apply to diverse BPS objects, such
as those dyons that appear in the generic strongly coupling
region of Seiberg-Witten theory.
In this paper, we wish to initiate a new framework that
can count and construct BPS states, without referring either to
specific subset of charges or to weak electric coupling,
but applicable to a large class of ${N}=2$ theories and BPS
states thereof.

One common lesson from  earlier studies of multi-centered BPS
states is that non-Abelian completion of the state at charge
centers is not essential for understanding wall-crossing,
since the latter is essentially a long distance
phenomenon from the spacetime viewpoint. The supergravity solutions
are all Abelian while, for solitons, it is the long range Coulomb-type
interactions that determined the multi-center nature of the state.
Related is the notion of the ``framed" BPS state \cite{Gaiotto:2010be}.
The main idea there was to treat one or more component dyons as an
external object, and the remainders as dynamical object around such
a background. This way, one can treat the former as the background,
in which the latter  moves around and sometime becomes supersymmetrically
bound to the core state. This split of the state into two parts can
simplify the state construction and counting substantially.

Inspired by these ideas, we wish to consider dyons moving
around purely Abelian dyonic background. In effect,
we will split the state in question into the heavy
``core state" of total charge $\g_c$ and light ``halo" or
``probe" of charge $\g_h$. For our purpose, it is the
ratio of the two masses that matters, so this can be
for instance achieved by approaching a singular point
where the probe dyon becomes massless. The low energy
dynamics of the probe dyon is quite natural
thing to do there since, precisely at such a singular point,
the probe dyon would be the lightest particle among charged
states. However, there are other circumstances where one part
become relatively light compared to the other, and our
framework will apply.

Another useful fact is that, as far as wall-crossing behavior goes,
we only need information near the relevant MSW's, away from which
the  BPS spectrum is continuous. This allows
another small quantity to play with, by taking vacuum very
near a marginal stability wall. As we will see later,
the distance to the MSW plays a role very similar to the weak
electric coupling in that it controls the nonrelativistic
approximation. In the end,
we find that the dynamics between the core and the probe
reduces to massive supersymmetric quantum mechanics with
two kinds of potentials.

These two lead us to a new model of low energy dynamics for
dyons in the strongly coupled
region of ${N}=2$ field theory. Although similar in spirit to the
old moduli dynamics of solitons, an essential difference here is
that the small electric coupling constant is no longer needed;
this is what allows us to apply the technique to much wider
class of BPS states than previously possible.\footnote{
In fact, it should be possible to extend this framework to
include gravity and discuss quantum bound states
of charged BPS black holes.}
The quantum mechanics has four supercharges, as required
by the BPS condition, but comes with only $3n$ bosonic
coordinates, three for each probe dyon, and $4n$ fermionic
coordinates. Compared to the conventional moduli dynamics
of weakly coupled regime, we are missing one angular collective
coordinate for each dyon.
This has something to do with the fact that we start with
dyons, rather than monopoles, as basic building blocks.

With this new low energy dynamics in place, we can
compute  how many BPS bound states of the
core and the probe dyons can form, and under what
condition. At the end of day we derive,
via a first-principle computation, the
semi-primitive wall-crossing
formula with $\g_1=\g_c$ and $\g_2=\g_h$. In this note
there is in fact no restriction on $\g_c$, as far as such
a state actually exist and all of its component dyon centers
can be made heavy.
Thus we in effect are computing  $\Omega(\g_c+n\g_h)$ with the
only restriction that the dyon $\g_h$ is primitive and become
massless somewhere in the vacuum moduli space. We wish to
emphasize that, alternatively,
we may think of the theory as a setup for finding framed
BPS state with line operator of charge $\g_c$ and halos $\g_h$ \cite{Gaiotto:2010be}.

This paper is organized as follows. In Section 2, we write
down the long-distance
Abelian form of the core state in terms of the central
charge function,  while the probe dyons are treated as
quantized solitons in that background. As a result we find
a bosonic low energy Lagrangian of the probe dyons purely
in terms of quantities that can be constructed out of the
central charge functions. This reproduces some of
general results, such as distances between two charge-centers,
obtained from supergravity attractor flow analysis, even though we
are dealing with field theory states.

Section 3 discusses how one can construct a $\CN=4$ supersymmetric
Lagrangian with $3n$ bosonic coordinates and $4n$ fermionic
coordinates, by extending previous studies by Coles and Papadopoulos
\cite{Coles:1990hr} and  also by \cite{Maloney:1999dv}.
These previous works constructed massless supersymmetric theories
of similar kind, which is, however, missing the crucial elements of
potentials. Without the latter, the bound states we are
interested in cannot form at all.
We construct in particular massive theories in which degrees of
freedom are cataloged by $SO(4)_R=SU(2)_L\times SU(2)_R$ algebra
with bosons in $({\bf 3},{\bf 1})$ (thus, the first $SU(2)_L$ also
serves as a rotation group) and fermions in $({\bf 2},{\bf 2})$
representations. The four supercharges are also in $({\bf 2},{\bf 2})$.
The Lagrangian has $SU(2)_R$ symmetry manifest while $SU(2)_L$
can be explicitly broken by the background.

Section 4 shows how the general discussion of section 3 makes contact
with the probe dyon dynamics of section 2 under the assumption
the vacuum moduli of the underlying ${N}=2$ theory is very near the MSW.
The latter assumption controls the energy scale of the potential
energy, and allows a nonrelativistic approximation possible.
We then quantize the resulting dynamics and derive the
bound states for $\g_c+\g_h$, and again shows how the
bound state size diverges as one approach MSW and
how the bound state is impossible on the other side of MSW.

Section 5 elevates this to a primitive wall-crossing formula,
and extends
further to the cases of $\g_c+n\g_h$ by invoking spin-statistics
theorem. This derives, in particular, the semi-primitive
wall-crossing formula from a first principle computation.

We then conclude in Section 6 with summary and other comments especially
on how one can make use of this formalism
to compute the line operator expectation values and how one
can extend the formalism to the supergravity setting. Some computational
details are summarized in Appendices.

\section{Classical Dynamics of Probe Dyons}

In this section, we construct the semiclassical form of the
core state, entirely in terms of the central charge function,
and describe energetics and dynamics of a probe dyon in the
core state background. This leads us to a bosonic Lagrangian
of the probe dyon, which will be supersymmetrized and quantized
in the later section.

Although the exercise here applies to any core state one can
imagine, as long as there solve the relevant semi-classical
BPS equation of the effective Abelian theory, we are eventually
interested in core states that actually exist as quantum BPS
states. It is known that the former does not always imply the
latter \cite{Mikhailov:1998bx,Gauntlett:1999vc}. Alternatively,
for the framed BPS states, the core state should correspond to
a supersymmetric line operator. Either way, we are interested
in case where the supersymmetric lift of this probe bosonic
dynamics would make sense in the context of the underlying
four-dimensional theory.

\subsection{Semiclassical Core State}

We start by recalling semiclassical properties of $N=2$ dyons
when expressed in terms of the low energy theory of Seiberg and
Witten. Traditionally the smooth solitons are possible only
when we include the entire non-Abelian origin, but this is
practical only in the weakly coupled limit.

To avoid such restrictions, a more convenient starting point
is to write the BPS equation in the Abelian low energy description
of Seiberg and Witten. This approach was investigated previously \cite{Argyres:2001pv,Ritz:2000xa}
with emphasis on split flow picture of the classical soliton
and gave an interesting parallel to the string web picture \cite{Bergman:1997yw} of
$N=4$ 1/4 BPS dyons. These solutions are invariably singular
at the charge centers, since there is no non-Abelain mechanism
to stop the Coulomb-like divergence at origin, which was
controlled ad hoc by introducing UV cutoffs.

For our purpose, however,
this divergence is of little consequence, essentially because
we will be using this solution as background. As long as we can ascertain
existence of quantum state of such a charge and as long as we
put correct boundary condition at such singular points, forcing
the probe dyon wavefunction to vanish there fast enough, there
would be no physical problem associated with it. It is entirely
analogous to the Hydrogen atom problem of undergraduate quantum
mechanics, where finite and trustworthy bound states are
obtained even though the Hamiltonian is naively singular at
origin.

Using the SUSY transformation rule for gaugino along
the particular direction parameterized by a phase factor $\zeta$,
one can obtain the BPS equations
\begin{eqnarray}\label{BPS1}
  \vec \CF_i - i \z^{-1} \vec \nabla \phi_i = 0 \ ,
\end{eqnarray}
where $i$ labels the unbroken $U(1)$ gauge groups, and
 $\vec\CF$ denotes the complexified field
strength 3-vector $ \vec B + i\vec E $. See appendix A for details.
There is also an electric version of this equation
\begin{eqnarray}\label{BPS2}
  \vec \CF_{D}^i - i \z^{-1} \vec \nabla \phi^i_D =0 \ ,
\end{eqnarray}
with $  \vec \CF_D^i \equiv \tau^{ij} \vec \CF_j$ and
\begin{equation}
\tau^{ij} = \frac{\partial^2}{\partial \phi_i \partial \phi_j} F_\text{SW}(\phi) \ ,
\end{equation}
where $F_\text{SW}(\phi)$ is the Seiberg-Witten prepotential of the given
theory.
Since it is $\text{Re}\CF_D$ that enters the Gauss constraint, the
field strengths are such that \cite{Argyres:2001pv}
\begin{eqnarray}
\text{Re}\,\int_{S^2_\infty}{\cal F}_i=4\pi P^i,\qquad
\text{Re}\,\int_{S^2_\infty}{\cal F}_D^i=-4\pi Q^i,
\end{eqnarray}
with the total magnetic charges $P^i$ and the total electric charges $Q^i$

In particular imagine a semiclassical core state of charges
$\g_c=(P^i,Q^i) =\sum_A \g_{c,A},$ with $\g_A=(P^i_A, Q^i_A)$,
distributed into several dyonic cores at $\vec x^A$, and the
field strength takes the following asymptotic forms,
\begin{eqnarray}
  &&\text{Re}\, \vec\CF^i = \sum_A \frac{P_A^i (\vec x-\vec x_A)}{|\vec x-\vec x_A|^3}
  =\vec\nabla\left(-\sum_A \frac{P_A^i }{|\vec x-\vec x_A|} \right) \ ,\nonumber\\
  &&\text{Re} \,\vec\CF_D^i = -\sum_A \frac{Q_A^i (\vec x-\vec x_A)}{|\vec x-\vec x_A|^3}
  =  \vec\nabla\left(\sum_A \frac{Q_A^i }{|\vec x-\vec x_A|}\right) \ .
\end{eqnarray}
One can show that $\z$ can be identified as the phase factor of
central charge $Z_\text{core}$ of this core state\footnote{See Appendix A.}
\begin{eqnarray}
  Z_\text{core}=|Z_\text{core}|\zeta = Q^i \phi_i (\infty)+  P_i  \phi_D^i (\infty)\ .
\end{eqnarray}
This semiclassical description is, strictly speaking, valid away from $\vec x=\vec x_A$'s.

Note that the positions, $\vec x_A$'s, of the centers would be restricted by
an analog of (\ref{denef}). Precise positions of these centers is, however,
immaterial for counting BPS bound states, as long as the relevant core state
actually exists as quantum and BPS bound state. This happens because one
ends up computing supersymmetric indices, which are robust under small deformations
of the supercharges. More important is how the core electromagnetic charge is
distributed into such centers. See section 4 for related discussions.

\subsection{Probe Dyons and Electromagnetic Forces}

Let us now introduce a probe particle of charge $\g_h=(p_i,q_i)$,
in a background created by such a core state. It will be
considered as a probe particle in the external electromagnetic field
by the massive core state. Using the equations (\ref{BPS1},\ref{BPS2}),
one obtains
\begin{eqnarray}\label{BPS3}
  q \cdot \vec \CF + p \cdot \vec \CF_D =i\zeta^{-1}  \vec \nabla {\cal Z}_h   \ ,
\end{eqnarray}
where ${\cal Z}_h=q \cdot \phi + p \cdot \phi_D$ is now understood as
position-dependent. We introduced the notation ${\cal Z}$ to emphasize
that this quantity is position-dependent. The usual central charge
${Z}$ is related to it as $Z={\cal Z}(\infty)$.

The real and imaginary part of the relation will give us hints how
to construct the low-energy Lagrangian of probe dyon in the
background of core particle. The real part can be succinctly written as
\begin{eqnarray}
  \vec \nabla V_\text{Coulomb}
  = -\vec \nabla \text{Re} \Big[ \zeta^{-1} {\cal Z}_h \Big]\ ,
\end{eqnarray}
where
\begin{eqnarray}
V_\text{Coulomb}&=&\text{Re}(\tau)_{ij}  \sum_A \frac{p^i P^j_A}{|\vec x-\vec x_A|} \nonumber\\
&& +\left(\text{Re}(\tau)\right)^{-1}_{ij}
  \sum_A \frac{\big( q_i + \text{Im}(\tau)_{ij} p^j \big)
  \big( Q_{j,A} +  \text{Im}(\tau)_{ij} P^j_A \big)}{|\vec x-\vec x_A|}
\end{eqnarray}
is nothing but the Coulomb potential energy felt by the probe dyon due to the core state.
The real part of this equation is even simpler
\begin{eqnarray}
  \vec \nabla \left(\sum_A \frac{Q_A\cdot p -P_A\cdot q}{|\vec x-\vec x_A|}\right)=
  -\vec \nabla \text{Im}\Big[\zeta^{-1} {\cal Z}_h \Big]\ ,
\end{eqnarray}
or equivalently
\begin{eqnarray} \label{imaginary}
  \vec \nabla \left( \sum_A \frac{\langle \g_{c,A}, \g_h \rangle}{|\vec x-\vec x_A|}\right) =
 - \vec \nabla \text{Im}\Big[ \zeta^{-1} {\cal Z}_h \Big]\ .
\end{eqnarray}
which, as we will presently see, encodes the Lorentz force on
the probe dyon.\footnote{The normalization of charges and the sign
convention for Schwinger product differs from that of Refs.~\cite{Denef:2000nb,Denef:2007vg}
$$\langle \g,\g'\rangle=\frac12 \langle \g',\g\rangle_\text{Denef-Moore} $$
Related is the fact that our $\zeta$ is $-\zeta_\text{Denef-Moore}$.}

We first discuss the invariant expression of the minimal coupling
under the Montonen-Olive duality.
Recall that, from the BPS equations, one can conclude that
$(\vec \CF, \vec \CF_D)$ transform under the duality transformation
as vector representation like $(\phi, \phi_D)$. For example, let us
consider the S-duality transformation of $(\vec \CF, \vec \CF_D)$
\begin{eqnarray}
 \vec B \ \to \ -\text{Im}(\tau) \vec E + \text{Re}(\tau) \vec B\ ,
  \qquad
\vec E \ \to \ \text{Im}(\tau)\vec B + \text{Re}(\tau) \vec E\ .
\end{eqnarray}
Then, one can easily show that, under the S-duality transformation,
\begin{eqnarray}
 q \ \to \ p\ , \qquad
 p \ \to \ - q\ ,
\end{eqnarray}
where we used, for the last transformation, the fact that $\t \to - \t^{-1}$.

When the probe dyon moves (slowly) under the electromagnetic field of
core particle, the minimal coupling terms therefore become \cite{Lee:1996kz,Gibbons:1995yw}
\begin{eqnarray}
  \CL_\text{int} =q^i \vec v \cdot \vec A^i
  + p^i \vec v \cdot \vec {\tilde A}^i
  + q^i A_0^i + p^i \tilde A_0^i\ ,
\end{eqnarray}
which is the duality invariant expression. Here
$A_\mu$ and $\tilde A_\mu$ are defined as
\begin{eqnarray}
  \text{Re} \vec \CF = \vec \nabla \times \vec A \ ,
  &&
  \text{Re} \vec \CF_D =\vec \nabla \times \vec{\tilde A}\ ,
  \nonumber \\
  \text{Im} \vec \CF = \vec \nabla \cdot A_0\ ,
  &&
  \text{Im} \vec \CF_D = \vec \nabla \cdot \tilde A_0\ .
\end{eqnarray}
Using the BPS equation (\ref{BPS3}), the interaction terms can
be managed into a rather simpler form
\begin{eqnarray}\label{int}
  \CL_\text{int} = -\vec v \cdot \vec{\cal W} + \text{Re}\Big[ \zeta^{-1} {\cal Z}_h(x)
  \Big] - \text{Re}\Big[ \zeta^{-1} {\cal Z}_h(\infty) \Big]\ ,
\end{eqnarray}
where the vector $\vec w$ satisfies the relation below
\begin{eqnarray}
  \vec \nabla \times \vec {\cal W} = \vec \nabla \text{Im}\Big[ \zeta^{-1} {\cal Z}_h(x) \Big]\ .
\end{eqnarray}
Note that, in (\ref{int}), $\text{Re}[\z^{-1} {\cal Z}_h(\infty)]=\text{Re}[\z^{-1} Z_h]$ represents
 the lowest possible energy the probe dyon can attain.

\subsection{Massive Moduli Dynamics of Probe Dyons}

Finally we come to the effect of the long range scalar field on the dyon.
The low-energy Lagrangian of  probe dyon
$\g_h$ moving in the background of core particle $\g_c$ can take the following
form
\begin{eqnarray}
  \CL^\text{bosonic} = \CL_\text{kin} + \CL_\text{int}\ ,
\end{eqnarray}
where the kinetic term must be \cite{Lee:1996kz,Gibbons:1995yw}
\begin{eqnarray}
  \CL_\text{kin} = - \big|{\cal Z}_h(x)\big| \sqrt{1-v^2} \simeq
  - |{\cal Z}_h(x)| + \frac12 \big|{\cal Z}_h(x)\big| \vec v^2 + \CO(v^4)
\end{eqnarray}
with ${\cal Z}_h(x) = q \cdot \phi + p \cdot \phi_D $, since $|{\cal Z}_h(x)|$
is the effective inertia of the probe dyon.
Adding all these together, we find the classical Lagrangian,
\begin{eqnarray}
  \CL^\text{bosonic} = \frac12 \big|{\cal Z}_h(x)\big| \vec v^2
  - \big|{\cal Z}_h(x)\big| + \text{Re}\left(\zeta^{-1} {\cal Z}_h(x)\right)
   - \text{Re}\Big[ \zeta^{-1} {\cal Z}_h(\infty) \Big]- \vec v \cdot \vec{\cal W}
\end{eqnarray}
with $ \vec \nabla \times \vec {\cal W} = \vec\nabla \text{Im}\left(\zeta^{-1}{\cal Z}_h(x)\right)$.

This Lagrangian has the classical ground state at $\vec x =\vec x_*$
where $\big|{\cal Z}_h(x_*)\big| = \text{Re}[\zeta^{-1} {\cal Z}_h(x_*)]$,
with the ground state energy $\text{Re}[ \zeta^{-1} {\cal Z}_h(\infty) ]$.
We wish to elevate this, later, to $\CN=4$ quantum mechanics, so it is more
convenient to separate out the ground state energy. Thus, our starting point is the
bosonic Lagrangian,
\begin{eqnarray}\label{bosonL}
  \CL_\text{moduli}^\text{bosonic} = \CL^\text{bosonic}+ \text{Re}\Big[ \zeta^{-1} {\cal Z}_h(\infty) \Big] \ ,
\end{eqnarray}
so that supersymmetric bound states would have zero energy.
This also reproduces an analog of Denef's formula  \cite{Denef:2000nb}
for the probe dyons since,
\begin{eqnarray}\label{distance}
\sum_A\frac{\langle \g_{c,A}, \g_h \rangle}{|\vec x_A-\vec x_*|}=\text{Im}\big[ \zeta^{-1} {\cal Z}_h(\infty) \big]\ .
\end{eqnarray}
from Eq.~(\ref{imaginary}) and $\text{Im} [\zeta^{-1}{\cal Z}_h(x_*)]=0$.
This is  the same as  (\ref{denef}) once we realize that
total central charge $Z_T=Z_c+Z_h$ is dominated by $Z_c$ since $Z_h/Z_c$ is
very small; $\zeta_T$ is approximately equal to $\zeta $.

\subsection{Fermionic Partners}

We have derived a classical (thus purely bosonic) Lagrangian that describe the dynamics of
a probe dyon in the background of the core state, with 3 bosonic collective
coordinates per each probe dyon. Without
much effort, we can further deduce that each probe dyon will also come with 4
fermionic degrees of freedom, giving $4n$ fermionic variables as
opposed to $3n$ bosonic variables.

The simplest way to see those four fermionic variables is to recall
that a BPS particle, of a given charge, in $N=2$ theory are at least
in the half-hypermultiplet, with spin content
\begin{equation}
[{ 1/2}] \oplus 2 [{ 0}]  \, .
\end{equation}
This spin content can be generated only if the dyon comes with
a pair of complex fermionic degrees of freedom in a spin 1/2 multiplet,
which translates to four real fermionic coordinates. They are, when
we consider the dyon in isolation, also Goldstino modes coming
from the four supercharges broken by the BPS state.
More generally,
the probe dyon could be in a BPS multiplet of type,
\begin{equation}
[s]\otimes\left([{ 1/2}] \oplus 2 [{ 0}] \right) \, ,
\end{equation}
with an angular momentum multiplet $[s]$ of spin $s$, in which
case $[s]$ typically arises because the probe dyon is itself
a composite or has, otherwise, some internal light degrees of
freedom. What matters for us is that we still have the same four
fermionic collective coordinates whose coupling to the bosonic ones
are tightly constrained by the $\CN=4$ supersymmetries.

When we consider the special limit of solitonic
dyons in weakly coupled theories, this mismatch between the
bosonic and the fermionic degrees of freedom can be understood
easily \cite{Weinberg:1979ma,Weinberg:1979zt,Weinberg:2006rq}.
Solitonic dyons arise there from excitation of a monopole
soliton with  particular $U(1)$ momenta turned on \cite{Julia:1975ff}. While
the initial monopole soliton comes with four bosonic and
four fermionic collective coordinates, one angular bosonic
coordinate is traded away in favor of its conjugate
momentum (which is physically the  electric charge). This procedure,
however, leaves the four fermionic coordinates intact. It has to be
so, since the dyon is still BPS and the necessary
half-hypermultiplet structure would be generated using all
four of these fermionic degrees of freedom.
Nor does this reduce the $\CN=4$ supersymmetry of the remaining dynamics,
although their embedding into the underlying field theory is
rotated in response to the new electromagnetic charges.

\section{Massive $\CN=4$ Mechanics onto Moduli Space}

An odd fact, when we consider a supersymmetric lift of the above
Lagrangian for probe dyons, is that the low energy
dynamics involves $3$ bosonic collective
coordinates for each probe dyon, yet, there should be 4 fermionic
counterparts. Supersymmetry with mismatching bosonic and fermionic
degrees of freedom is in principle possible for quantum mechanics
because there is no notion of spin, but still construction of
such theories, especially with extended supersymmetry, was not
widely studied. The only known example is certain (massless) class
of supersymmetric nonlinear sigma models by Coles et. al. \cite{Coles:1990hr},
which were later specialized in the context of extremely charged black
holes of the same charges \cite{Maloney:1999dv}. Neither of these
studies considered massive versions, as needed here, however.

Similar situation existed a dozen years ago
when low energy dynamics of solitonic monopoles were studied for weakly
coupled ${N}=2,4$ Yang-Mills theories. The conventional
massless moduli dynamics \cite{Manton:1977er,Manton:hs,Atiyah-Hitchin}
with $4n$-dimensional target manifolds without potential
were found to be inadequate  for  dyons in generic
Coulombic vacuum when the rank of the gauge group is two or
larger \cite{Lee:1998nv}. The problem was the lack of
potential terms in this older formulation. The low energy
dynamics of monopoles had to be reformulated
so that both the potentials and $\CN=4$ supersymmetry are
manifest. Later, such massive $\CN=4$  quantum mechanics were found,
simply by twisting supercharged by triholomorphic Killing vector
fields on the moduli space
\cite{Bak:1999da,Bak:1999vd,Gauntlett:1999vc,Gauntlett:2000ks}.\footnote{Some
related mathematical structures were first
studied in Refs.~\cite{AlvarezGaume:1983ab} while its potential
connection to dyons was previously hinted by Ref.~\cite{Tong:1999mg}.}
This lead to a whole machinery whereby dyon spectra in the
weakly coupled limit of $N=2,4$ Yang-Mills theories were
constructed explicitly \cite{Stern:2000ie}. See Ref.~\cite{Weinberg:2006rq}
for a broad overview of this development.

In this section, we wish to investigate how the new kind of classical
low energy dynamics of section 2 can be also elevated to one with
$\CN=4$ supersymmetry.
We will find that massive $\CN=4$  supersymmetric mechanics with
mismatching bosonic and fermionic degrees of freedom is possible
and will, specifically, build  a massive (i.e. with potential) supersymmetric
Lagrangian with 3 bosonic coordinates and 4 fermionic
coordinates. This restriction to the lowest possible target
dimension simplifies the construction greatly, in part because the target
manifold turned out to be conformally flat $R^3$, and yet still good
enough for deriving semi-primitive wall-crossing formula.\footnote{
For generalization that can address many probe
dyons with non-negligible mutual interactions,
we need to consider higher dimensional target manifolds, which
is left for a future work.}

\subsection{Toy Model: Flat $R^3$ Target}

As a toy model, let us pretend that the bosonic moduli space is
flat $R^3$ and see how scalar and vector potentials on $R^3$ can be
incorporated into the quantum mechanics in a manner consistent
with four supercharges.


$\CN=1$ supersymmetry is easy to incorporate. We start with the usual
transformation rule
\begin{align}
  \d x^a = - i \e \psi^a \ , \qquad
  \d \psi^a = \e {\dot x}^a \ , \qquad
\end{align}
under which the following free Lagrangian that is invariant
\begin{align}\label{free}
  \CL^{(0)} = \frac12 {\dot x}^a {\dot x}^a + \frac{i}{2} \psi^a {\dot \psi}^a \ .
\end{align}
Since we are dealing with quantum mechanics, rather than
a field theory, we can add any number of fermions, as long as we
let them be invariant under the above supersymmetry transformation.
As we will see shortly, however, extended supersymmetry would not
leave this extra fermion intact.

For our purpose, one extra fermion $\lambda$ is needed
for each triplet of $(x^a,\psi^a)$, so we may start with
\begin{align}
  \CL^{(0)} = \frac12 {\dot x}^a {\dot x}^a
  + \frac{i}{2} \psi^a {\dot \psi}^a+ \frac{i}{2} \lambda {\dot \lambda}\ ,
\end{align}
where
\begin{align}
  \d \lambda=0 \ .
\end{align}
Incorporation of  an external gauge field $w$ on $R^3$ is equally
easy. Adding a minimal coupling $-\dot x^aw_a$ to the Lagrangian and
noting the supersymmetry transformation property,
\begin{align}
  \d \big( -  w_a {\dot x}^a \big) = & + i \e \psi^a {\dot x}^b \big(
  \partial_a w_b - \partial_b w_a \big) + \text{total derivative} \
  \nonumber \\
  = & +i  \partial_a w_b \big( \e \psi^a {\dot x}^b - \e \psi^b {\dot x}^a \big)\ ,
\end{align}
one finds a canceling term of type
\begin{align}
  \d \big( + i \partial_a w_b \psi^a \psi^b \big)
  =  + i \partial_a w_b \big( {\dot x}^a \e \psi^b - {\dot x}^b \e \psi^a \big)\ .
\end{align}
In summary, the following Lagrangian has ${\cal N}=1$ supersymmetry
\begin{align}
  \CL = \frac12 {\dot x}^a {\dot x}^a + \frac{i}{2} \psi^a {\dot \psi}^a+ \frac{i}{2} \lambda {\dot \lambda} - w_a {\dot x}^a
  + i \partial_a w_b \psi^a \psi^b\ .
\end{align}
%
In order to introduce the bosonic potential to the above model, we
modify the transformation rule of the auxiliary fermion $\lambda$ as
\begin{align}\label{n=1}
  \d x^a = - i \e \psi^a \ , \qquad
  \d \psi^a = \e {\dot x}^a \ , \qquad
  \d \lambda = \e K\ ,
\end{align}
upon which we find
\begin{align}
  \d \left(\frac12 {\dot x}^a {\dot x}^a + \frac i2 \psi^a {\dot \psi}^a
  + \frac i2 \lambda {\dot \lambda} - w_a {\dot x}^a + i \partial_a w_b
  \psi^a \psi^b \right)= - i \e \l {\dot x}^a \partial_a K \ .
\end{align}
The canceling term for this is
\begin{align}
  \d \big( +i \partial_a K \psi^a \l \big) = &
  i \e \lambda {\dot x}^a \partial_a K - i \e \psi^a K
  \partial_a K \nonumber \\
  = &  i \e \lambda {\dot x}^a \partial_a K + \d x^a  K  \partial_a K\ ,
\end{align}
while one must add one more to close the transformation algebra,
\begin{eqnarray}
  \d \big( -\frac12 K^2 \big) = - \d x^a K \partial_a K\ .
\end{eqnarray}
In summary, the Lagrangian
\begin{align}\label{N=1Lagrangian}
  \CL= \frac12 {\dot x}^a {\dot x}^a + \frac i2 \psi^a {\dot \psi}^a
  + \frac i2 \lambda {\dot \lambda} - w_a {\dot x}^a  - \frac 12 K^2
  + i \partial_a w_b \psi^a \psi^b + i \partial_a K \psi^a \l\
\end{align}
has $\CN=1$ supersymmetry, for any $K$ and $w$.


We eventually wish to formulate dyon dynamics with
$\CN=4$ supersymmetries. For conventional supersymmetric quantum
mechanics, this requires the target manifold to be $4n$ dimensional and
hyperK\"{a}hler, which is clearly inappropriate for our $3n$
dimensional target. Nevertheless, the BPS nature of the dyons and
existence of BPS bound states implies that there should exist such an $\CN=4$ lift.

To find the relevant supersymmetries and the subsequent restrictions
on the potentials, note that, since the number of bosons and
the number of fermions mismatch by 3 to 4, we can organize the
degrees of freedom using $SO(4)=SU(2)_L\times SU(2)_R$ algebra.
Let the bosons, $x^a, a=1,2,3$, transform as $({\bf 3,1})$ representation
while the fermions, $(\psi^a,\lambda)$, are naturally in $({\bf 2,2})$
and better denoted as $\psi^m, m=1,2,3,4$ with $\psi^4=\lambda$.
Thus, $a,b,\dots$ are the vector indices of $SU(2)_L$ while $m,n,\dots$
are vector indices of $SO(4)$. The $\CN=4$ supersymmetries are
then naturally in $({\bf 2,2})$ under this $SO(4)$, since it should
relate bosons to fermions. Thus the four supersymmetry transformation
parameters will be denoted by $\e_m$.

A useful method of relating $SU(2)_L$ objects to $SO(4)$ object is
to employ 't Hooft's self-dual symbol $\eta^a_{mn}$. Based on
previous experience of embedding $SO(3)\simeq SU(2)_L$ into $SO(4)$, such as
in Yang-Mills instanton construction,   one can guess the
following $\CN=4$ SUSY transformation rules
\begin{align}\label{SUSYrule}
  \d x^a = i \eta^a_{mn} \e^m \psi^n\ ,
  \qquad
  \d \psi_m = \eta^a_{mn} \e^n {\dot x}^a
  + \e_m K\ ,
\end{align}
with the 't Hooft self-dual symbol $\eta$ defined as \cite{'tHooft:1976fv}
\begin{equation}
\eta^a_{bc}=\epsilon_{abc},\qquad \eta^a_{b4}=\delta^a_b=- \eta^a_{4b} \ .
\end{equation}
which, for $\e^4\equiv \e$, matches (\ref{n=1}).

This suggests that the Lagrangian (\ref{N=1Lagrangian}) can be extended to
admit $\CN=4$ supersymmetries, if we can organize the fermion bilinears in
terms of $\eta$ symbol as
\begin{align}
  \CL=  \frac12 \sum_{a=1}^3 {\dot x}^a {\dot x}^a
  + \frac i2 \sum_{m=1}^4 \psi^m {\dot \psi}^m
  + \frac i2 \eta^a_{mn} \partial_a K \psi^m \psi^n
  - w_a {\dot x}^a - \frac12 K^2\ ,
\end{align}
which matches (\ref{N=1Lagrangian})  if we impose
\begin{align}
  \e^{abc} \partial_a K = \partial_b w_c - \partial_c w_b\ .
\end{align}
One can indeed show that the above Lagrangian is invariant under
the $\CN=4$ SUSY transformation rules (\ref{SUSYrule}). This
Lagrangian is manifestly invariant under $SU(2)_R$. The $SU(2)_L$
invariance is broken only to the extent that $K$ breaks the
rotational invariance. If $K$ is spherically symmetric, for instance,
the full $SO(4)$ symmetry would be restored.

Let us discuss the closure of the $\CN=4$ algebra.
For bosonic variables, one can show
\begin{align}\label{n=4flat}
  \d_\z \d_\e x^a = - i \eta^a_{mn} \eta^b_{pn} \e^m \z^p {\dot x}^b
  + i \eta^a_{mn} \e^m \z^n K \ ,
\end{align}
which implies
\begin{align}
  \big( \d_\z \d_\e - \d_\e \d_\z \big) x^a = - 2i \e^m \z^m {\dot x}^a \ .
\end{align}
Here we used the following identity $\eta^a_{mn}\eta^b_{pn} = \d^{ab} \d_{mp} + \e^{abc}
\eta^c_{mp}$. Let us now in turn consider the case of fermionic variables.
\begin{align}
  \big( \d_\z \d_\e - \d_\e \d_\z \big) \psi_m = &
  - 2 i \e^n \z^n {\dot \psi}_m +
  i \big( \e^n \z^m + \e^m \z^n \big) {\dot \psi}^n
  + i \eta^a_{pq} \partial_a K \big( \e^p \z^m + \e^m \z^p \big) \psi^q\ ,
  \nonumber \\
  = & - 2 i \e^n \z^n {\dot \psi}_m\ ,
\end{align}
where for the last equality we used  the equation of motion of $\psi^m$
\begin{align}
  {\dot \psi}^q + \eta^a_{qn} \partial_a K \psi^n = 0\ .
\end{align}
We therefore conclude that the ${\cal N}=4$ SUSY algebra is given by
\begin{align}
  \big\{ Q_m , Q_n \big\} = 2 \d_{mn} H\ ,
\end{align}
with the Hamiltonian $H$.

\subsection{Toy Model with $\CN=1$ Superfields }

We can write the above Lagrangian by
introducing $\CN=1$ superspace with an anti-commuting
coordinate $\theta$.
Following the notation in Ref.~\cite{Maloney:1999dv},
we define the bosonic and the fermionic superfields as
\begin{equation}
\Phi^a=x^a-i\theta\psi^a,\quad \Lambda=i\lambda+i\theta b \ .
\end{equation}
The supersymmetry
generator and the supercovariant derivatives are then,
\begin{equation}
Q=\partial_\theta+i\theta\partial_t,\quad D=\partial_\theta -i\theta\partial_t \ .
\end{equation}
Our toy model based on flat $R^3$, with scalar and
vector potentials, can be written in a superspace form
as
\begin{equation}
 {\cal L}=\int  d\theta \;\left(\frac{i}{2}D\Phi^a\partial_t\Phi^a -\frac12\Lambda D\Lambda
+iK(\Phi)\Lambda -i w(\Phi)_a D\Phi^a\right) \ .
\end{equation}
Although only $\CN=1$ supersymmetry is manifest, we saw
that $\CN=4$ supersymmetry will emerge if the condition $* dK= dw$ is
imposed.
This form of the Lagrangian is useful because it could be
generalized to the curved moduli space almost immediately.

\subsection{Massive $\CN=4$ Theory onto Conformally Flat $R^3$ }

Recall that, for a single probe dyon, there are three quantities that
appears in the bosonic moduli dynamics. The scalar and the vector
potentials, as we already incorporated into $\CN=4$ toy model above,
and most crucially, the position-dependent mass term $|{\cal Z}_h|$ for
the coordinates $x^a$. Thus, in addition to the above interaction
terms, we wish to replace $R^3$ by a conformal flat $R^3$ whose metric is
\begin{equation}
g_{ab}=f\delta_{ab} \ ,
\end{equation}
with $f$ later to be identified with $|{\cal Z}_h|$.
In fact, as can be inferred from the massless version in Refs.~\cite{Coles:1990hr,Maloney:1999dv},
$\CN=4$ supersymmetry restricts the three-dimensional metric to be conformally flat.

We defer  detailed construction 
to appendix B, and  simply state here  that the desired Lagrangian, now with
potentials, has the superspace form
\begin{eqnarray}\label{super}
 {\cal L}&=&\int  d\theta \;\biggl(\frac{i}{2}f(\Phi)
D\Phi^a\partial_t\Phi^a -\frac12f(\Phi)\Lambda D\Lambda \nonumber\\
&&+\frac14\epsilon_{abc}\partial_af(\Phi)D\Phi^bD\Phi^c\Lambda
 +i{\cal K}(\Phi)\Lambda -i {\cal W}(\Phi)_a D\Phi^a\biggr) \ ,
\end{eqnarray}
with the condition  
\begin{equation}\label{n=4condition}
\partial_a{\cal K}=\e_{abc}\,\partial_b{\cal W}_c
\end{equation}
imposed. In terms of component fields, this equals
\begin{eqnarray}\label{n=4}
  \CL &=
  & \frac12 f \left( {\dot x}^a {\dot x}^a  + i \psi^m \nabla_t \psi^m\right) \nonumber\\
  &&
  - \frac{1}{4\cdot 4!}   \left( 2\partial^2 f - f^{-1}(\partial  f)^2 \right)
  \e_{mnpq}\psi^m \psi^n \psi^p \psi^q
  \nonumber \\ &&
  - \frac{1}{2 f} {\cal K}^2 - {\cal W}_a {\dot x}^a
  + \frac i2 f^{1/2} \partial_a \big( f^{-1/2} {\cal K} \big) \eta^a_{mn} \psi^m \psi^n \ ,
\end{eqnarray}
where  the covariant derivative for fermions is given by
\begin{align}
  \nabla_t \psi^m = {\dot \psi}^m
  + \frac12 \e_{abc} {\dot x}^a f^{-1} \partial_b f \eta^c_{mn} \psi^n\ .
\end{align}
As in the flat case, the degrees of freedom and the supercharges
are cataloged by  $SO(4)=SU(2)_L\times SU(2)_R$ algebra, and the
Lagrangian is manifestly invariant under $SU(2)_R$. The $SU(2)_L$
keeps track of how $f$ and ${\cal K}$ (and thus $\CW$ also) transform under spatial rotations,
and become a symmetry whenever these quantities are rotationally invariant.

This $SO(4)$ structure and $SU(2)_R$ symmetry tells us an extended $\CN=4$
supersymmetry exists, as in the flat $R^3$ example. It is not difficult to see
that
\begin{eqnarray} \label{susy-off}
  \d_\e x^a&=&  i \eta^a_{mn} \e^m \psi^n \ , \qquad \nonumber\\
  \d_\e \psi_m &=& \eta^a_{mn} \e^n {\dot x}^a + \e_m b \; ,
\end{eqnarray}
with four Grassman parameters $\e^m$ leaves the Lagrangian invariant.
The only difference
from the flat case, (\ref{n=4flat}), is that $K$ is replaced by its generalized
form, namely on-shell value of
the ${\cal N}=1$ auxiliary field $b$,
\begin{eqnarray}
b=\frac{1}{f}\,\left({\cal K} +\frac i4 \eta^a_{pq}
  \partial_a f \psi^p \psi^q \right) \ .
\end{eqnarray}
The superalgebra remains the same as the flat case,
\begin{align}
  \big\{ Q_m , Q_n \big\} = 2 \delta_{mn} H\ ,
\end{align}
where we denoted the four supercharges by $Q_m$ as before
and the Hamiltonian by $H$. For completeness, we also record
the classical form of the Hamiltonian,
\begin{eqnarray}
  H_{classical} &=
  & \frac{1}{2f}   \pi^a \pi^a   + \frac{1}{4\cdot 4!}\left(2 \partial^2 f - f^{-1} (\partial f)^2 \right) \e_{mnpq}
  \psi^m \psi^n \psi^p \psi^q   \nonumber \\
  &&+ \frac{1}{2 f}\, {\cal K}^2
   - \frac i2 f^{1/2} \partial_a \big( f^{-1/2} {\cal K} \big) \eta^a_{mn}
  \psi^m \psi^n
  \ ,
\end{eqnarray}
with the covariantized momenta
\begin{equation}
\pi^a=p_a+{\cal W}_a -\frac{i}{4}\epsilon_{abc}\partial_bf\eta^c_{mn}\psi^m\psi^n  \ .
\end{equation}
The quantum Hamiltonian differs from this by normal ordering issue,
and can also be found in appendix B.

\section{Quantum BPS States near WMS}

\subsection{$\CN=4$ Low Energy Dynamics of Dyons near MSW}

Let us stop here and ask under what circumstances we actually expect
to see a sensible low energy dynamics of dyons to appear. The old
setting based on dyons as quantum bound states of excited magnetic
solitons was possible by resorting to weakly coupled regime. There,
the basic requirements was that the energy due to electric charges
and also due to  motion of the solitons are of higher order.
Thus, the reduction to quantum mechanics is
controlled two small parameters; typical speed of the magnetic
soliton and the electric coupling constant \cite{Manton:1977er}.

Here, however, we are here dealing with dyons of generic charges
at generic coupling, and must find  different criteria to justify
reduction to low energy quantum mechanics. Note that the weak coupling requirement
and the small speed requirement of old moduli dynamics is in fact
interrelated. That happens was that the moduli dynamics of ${N}=2$
and ${N}=4$ monopoles usually acquire a bosonic potential
of order $e^2$, so for typical states the small electric coupling
is necessary to ensure small velocities.

In the present low energy dynamics of probe dyons around a core
state, the size of the potential is instead controlled by how far
are the phases of central charges of core and halo particles are
aligned. Thus, by staying very near MSW, we have
a good control over the potentials.
Furthermore, the massgap between this
sector and the rest is also substantial, and controls possible
interference from other charged particles.\footnote{The latter is easiest
to see when the small mass ratio is achieved by being near a
singular point of the vacuum moduli space. The relevant coupling that
governs the interaction of the field theory would be a dualized
coupling which becomes small as the singular point is
approached.} So it is the
proximity to the MSW and also the mass ratio of the two parts
that now control the reduction to the low energy quantum mechanics.

With this mind, we compare (\ref{bosonL}) against the
supersymmetric Lagrangian (\ref{n=4}). One can see the supersymmetric
uplift may work only if
\begin{align}
  f = |{\cal Z}_h|\ , \qquad \frac{1}{2f}{\cal K}^2 = |{\cal Z}_h|-{\rm Re}[\zeta^{-1}{\cal Z}_h],  \qquad
  \vec \nabla \times \vec {\cal W}=\vec\nabla \left( \text{Im}[\zeta^{-1}{\cal Z}_h]\right) \ .
\end{align}
but the requisite $\CN=4$ relationship between $\CK$ and ${\cal W}$, $*d\CK=d\CW$,  is
not yet apparent. Thankfully, this condition is satisfied precisely when
the criteria for the low energy approximation are met, as we described
above.

To see the latter, write $\zeta^{-1}{\cal Z}_h =|\,{\cal Z}_h|\,e^{i\beta}$.
Near the wall of marginal stability, the angle $\beta$ at spatial infinity
is very small whereas its value at classical vacuum is 0. Recall that the
bound states we wish to find and count are all peaked at the classical vacuum
manifold. This allows us to expand relevant quantities in small $\beta$.
As we move closer to charge centers,  $\vec x_A$'s, $\beta$ can grow again
but the precisely form of the background at such
charge centers are not to be trusted and also happily immaterial for our
purpose of finding BPS bound states.  Therefore, we take the value of
$\beta$ to be small  everywhere and find
\begin{equation}
{\cal K}^2 =  2|\,{\cal Z}_h|^2(1-\cos\beta)\simeq |\,{\cal Z}_h|^2\beta^2\simeq
|\,{\cal Z}_h|^2(\sin\beta)^2=\left(\text{Im}[ \zeta^{-1}{\cal Z}_h]\right)^2 \ .
\end{equation}
Thus, for all practical purpose,  we may identify ${\cal K}=\text{Im}[\zeta^{-1}{\cal Z}_h]$
and the $\CN=4$ requirement (\ref{n=4condition}) is obeyed automatically. This completes
the derivation of $\CN=4$ moduli dynamics in (\ref{n=4}) of a  probe dyon
in a given core state background.

The function ${\cal K}$ can be generally written, from (\ref{imaginary}), as
\begin{equation}
{\cal K}= {\cal K}_0-\sum_A\frac{\langle \g_{c,A},\g_h\rangle }{|\,\vec x-\vec x_A|} \ ,
\end{equation}
with $\g_{c,A}$ centers of the core states at $\vec x_A$ and also
${\cal K}_0\equiv \text{Im}[\z^{-1}{\cal Z}_h(\infty)].$ Details of $f=|{\cal Z}_h|$
won't matter much for the purpose of constructing bound states, it turns out,
as long as we keep track of its singular behaviors at charge centers.

Before we start the detailed analysis, let us note again that the semiclassical
core state here is not really a good representation very near its charge
center(s), where the non-Abelian nature of the states becomes relevant.
Naturally, the low energy dynamics of probe
dyons is  plagued by the same issue. However, this hardly matters near MSW
because the bound state (if any) would be very large and determined entirely
by Abelian part of the low energy field theory: Whatever singularity at
Coulombic centers cannot alter such wavefunction significantly, as long as we impose
the boundary condition at centers intelligently enough. This should become
more obvious when we discuss actual bound state wavefunctions in section
4.3. For this reason, we may as well take the above form of ${\cal K}$, $f$, etc
literally, and consider supersymmetric bound states thereof, with
some care given to the boundary condition of the wavefunctions at centers $\vec x_A$.

\subsection{Quantization and Supercharges}

Let us start with the canonical commutators. The conjugate
momenta  of bosons are denoted as $p_a$,
\begin{equation}
[p_a,x^b]=-i\delta^b_a \ ,
\end{equation}
whereas the normalized fermions,
$\hat\psi^m\equiv {f}^{1/2}\psi^m$, are more convenient for
writing out the remaining
canonical commutators,
\begin{equation}
\{\hat\psi^m,\hat\psi^n\}=\delta^{mn}, \qquad [p_a,\hat\psi^m]=0=[x^a,\hat\psi^n] \ .
\end{equation}
With this we can now write the four supercharges as
\begin{align}
  Q_m = - \eta^a_{mn} \psi^n (p_a + \CW_a)
  + \frac i4  \eta^a_{mn} f^{-1} \partial_a f \psi^n +
  \frac i4 \partial_a f \eta^a_{pq}
  \psi^{[p} \psi^q \psi^{m]} +  \CK \psi^m \ .
\end{align}
For  the proof that these are right supercharges,
see appendix B.
In particular, the supercharge associated with $\epsilon^4$ is
\begin{equation}
Q=Q_4=\psi^a\left(p_a+{\cal W}_a\right)-\frac{i}{4}f^{-1}\partial_af\psi^a
+\frac{i}{4}\partial_a f\e_{abc}\psi^b\psi^c\lambda
+\lambda{\cal K} \ .
\end{equation}
Since the superalgebra implies $\{Q_m,Q_n\}=2\delta_{mn} H$, the ground
state of the system can be found by demanding that it be
annihilated by $Q_4$.

\subsection{BPS Bound States and Marginal Stability}

The canonical commutator of the fermions
\begin{equation}
\{\hat\psi^m,\hat\psi^n\}=\delta^{mn}
\end{equation}
is a Clifford algebra which can be represented by Dirac matrices,
\begin{equation}
\sqrt{2}\;\hat\psi^a=\gamma^a=\left(
\begin{array}{ll}
0 & \sigma^a \\
\sigma^a & 0
\end{array}
\right),\qquad
\sqrt{2}\hat\lambda=\sqrt{2}\;\hat\psi^4=\gamma^4=\left(
\begin{array}{cc}
0 & i \\
-i & 0
\end{array}
\right) \ ,
\end{equation}
and wavefunctions can be regarded as 4-component spinors on $R^3$.
Also useful is the chirality operator
\begin{equation}
\Gamma\equiv \gamma^1\gamma^2\gamma^3\gamma^4= \left(
\begin{array}{cc}
1 & 0 \\
0 & -1
\end{array}
\right) \ .
\end{equation}
Under the above representation, one of supercharge $Q_4$
now has a simple form,
\begin{equation}
\sqrt{2f}\; Q_4=
\g^a (p_a + \CW_a) - \frac i2(\partial_a \text{log}f) \g^a\, \frac{1 - \Gamma}{2} + \CK \g^4 \ ,
\end{equation}
or more explicitly,
\begin{equation}
\sqrt{2f}\; Q_4=\left(\begin{array}{cc}
0 & \sigma\cdot(p+{\cal W}) +  i{\cal K} \\
\sigma \cdot(p+{\cal W})  - i  \s \cdot \partial (\log f^{1/2}) -  i{\cal K} & 0
\end{array}\right) \ .
\end{equation}
We wish to find supersymmetric ground states, $Q_4\Psi=0$. Since
${\cal H}\Psi=0$ then, such states would actually preserve all four supercharges.
Such states are then automatically BPS with respect to the $N=2$ field theory
with the energy $\text{Re}[ \zeta^{-1} { Z}_h(\infty) ]$, as can be seen from (\ref{bosonL}),
not counting the  core state energy.

Write the four-component wavefunction $\Psi$ as
\begin{equation}
\Psi=\left(
\begin{array}{c}
f^{-1/2} {\cal U} \\
{\cal V}
\end{array}
\right) \ ,
\end{equation}
upon which two component wavefunctions ${\cal U}$ and ${\cal V}$ obey
\begin{eqnarray}
\left(\sigma\cdot(p+{\cal W})-i{\cal K}\right){\cal U}&=&0,\nonumber\\
\left(\sigma\cdot(p+{\cal W})+i{\cal K}\right){\cal V}&=&0,
\end{eqnarray}
With the supersymmetry condition $d{\cal K}=*d{\cal W}$, it is
easy to see that the first equation cannot have a normalizable
solution while the second may. Denoting the respective operators as
${\cal D}_\pm$,
\begin{equation}
{\cal D}_\mp{\cal D}_\pm=\left(p+{\cal W}\right)^2 +{\cal K}^2
+\sigma^a\left(\partial_a{\cal K}\pm\partial_a{\cal K}\right) \ ,
\end{equation}
which shows that ${\cal D}_+{\cal D}_-$ is a positive definite operator
while ${\cal D}_-{\cal D}_+$ is not. Only the latter can have zero modes.
Thus, we arrived at the conclusion
that the counting of BPS bound states between the core dyon and the
probe dyon becomes that of counting normalizable two-component zero
modes ${\cal V}$ of the operator ${\cal D}_+$, with the final form of
the BPS bound state
\begin{equation}
\Psi=\left(
\begin{array}{c}
0\\
\CV
\end{array}
\right) \ ,
\end{equation}
with ${\cal D}_+\CV=0$.

It is illuminating to solve this equation for the particular case of
spherically symmetry core state. The vector potential would be that
of a Dirac monopole, so we denote
\begin{equation}
{\cal W}= -g A_\text{Dirac},\qquad g=-\langle \g_c,\g_h\rangle, \quad A_{Dirac}=-\cos\theta d\phi \ ,
\end{equation}
from which follows the scalar potential
\begin{equation}
{\cal K}= {\cal K}_0+\frac{g}{r} \ .
\end{equation}
In this case $SU(2)_L$ also becomes a symmetry, allowing
an explicit solution to the bound state problem.
The number $g$ is half-integer quantized, as dictated by the
Dirac quantization of this quantum mechanics, and also from
its field theory origin as the Schwinger product of the two
quantized charge vectors. The angular and spin part of the
wavefunction is classified by spinorial monopole spherical
harmonic tensor. The lowest possible angular momentum would
be than $j=|g|-1/2$, since the charge interacting with such
a Dirac monopole, ${\cal W}$, is endowed with a well-known
angular momentum $-g\hat r$. Tensoring with the intrinsic
spin 1/2, the minimum possible value $j=|g|-1/2$ follows.

Denoting the corresponding the lowest-lying
two-component angular momentum eigen-states $\eta_{j=|g|-1/2,m}$ of $SO(3)\simeq SU(2)_L$
we rely on  Kazama et.al. \cite{Kazama:1976fm} for reduction of the above
to the radial equation,
\begin{equation}
{\cal V}=h(r)\eta_{j=|g|-1/2,m},\qquad
\left(-i\frac{g}{|\,g|}\times\left[\frac{d}{dr}+\frac{1}{r}\right]+i{\cal K}(r)\right)h(r)=0 \ .
\end{equation}
Integrating the latter equation, we find
\begin{equation}
h(r)=\frac{1}{r}\exp\left( \frac{g}{|g|}\int^r {\cal K}(r)\right)
=C\,r^{|\langle \g_c,\g_h\rangle|-1}
\exp\left(-[\text{sgn}(\langle \g_c,\g_h\rangle)\cdot{\cal K}_0] \cdot r\right) \ ,
\end{equation}
with the normalization constant $C$.
Note that this gives a normalizable  ground state
if and only if the half-integer-quantized
$ \langle \g_c,\g_h\rangle$ is not zero
and has the same sign as ${\cal K}_0= {\text{Im}[\zeta^{-1}{\cal Z}_h(\infty)]}$.\footnote{$L^2$
normalizability requirement from
$r=0$ region is satisfied as long as $|\langle g_c,\g_h\rangle|$ is
not zero, so does not impose additional restriction.}
The latter condition is also reflected on the fact that
the probability density of this wavefunction is peaked at radial size
\begin{equation}
\frac{\langle\g_c,\g_h\rangle}{\text{Im}[\zeta^{-1}{\cal Z}_h(\infty)]} \ ,
\end{equation}
which, for a single-center core state, exactly mirrors the classical orbit radius in (\ref{distance}).

The sign of $ \langle \g_c,\g_h\rangle$ is determined
by the charges of the core state and the probe state, and does not
change as we move along the vacuum moduli space. However, the
 ${\cal K}_0=\text{Im}[\zeta^{-1}{\cal Z}_h(\infty)]$
does change its sign across the marginal stability wall
between the core state and the probe state. Classically,
this happens because $g{\cal K}_0<0$ would make the
potential repulsive. The upshot is that
the BPS bound states of one side where $\langle \g_c,\g_h\rangle/ \text{Im}[\zeta^{-1}{\cal Z}_h(\infty)]>0$
disappear as we move to the other side where $\langle \g_c,\g_h\rangle/\text{Im}[\zeta^{-1}{\cal Z}_h(\infty)]<0$,
as was originally found in the supergravity setting.

With this exercise, we learned a few things:
\begin{enumerate}
  \item[$\bullet$] Normalizable bound state between the core state
  and the probe state is realized only when the Schwinger product
  of the two charge is nonzero.

  \item[$\bullet$] Normalizable bound state between the core state
  and the probe state is realized only when the Schwinger product
  of the two charge is of the same sign relative to the value of
  $\text{Im}[\zeta^{-1}{\cal Z}_h]$ at spatial infinity.

  \item[$\bullet$] When such normalizable states exist, the degeneracy
  is $2j+1=2|\langle \g_c,\g_h\rangle|$.

\end{enumerate}
Much of the above statements are properties of a Dirac operator
with ${\cal D}_\pm$  as the chiral and the anti-chiral
parts; there must be an index theorem associated
with them.

In fact, the structure of the operators are essentially
that of an electrically charged fermionic field around the magnetic monopole,
except that we do not see the
non-Abelian structure that regulate the short-distance behavior of the core state.
Similar issues in the context of quantization in the backgrounds
of non-Abelian monopoles vs. Dirac monopoles (or more precisely Wu-Yang monopoles \cite{Wu:1976ge})
have been studied in
depth decades ago, where it was found that with proper boundary
condition at origins of the latter, behaviors of the two are essentially
the same \cite{Yamagishi:1982wp}. The boundary condition is constrained by the requirement
that the Dirac operator constructed out of ${\cal D}_\pm$ should be
Hermitian, which is known in the literature as the self-adjoint extension.

This is related to the fact that, even though the two
potentials of the quantum mechanics are singular at origin,
the wavefunctions found are regular everywhere
and in particular suppressed strongly at origin. If we attempted to
solve for ${\cal D}_-{\cal U}=0$, the radial eigen-function of ${\cal U}$ would have
the behavior $r^{-|\langle \g_c,\g_h\rangle|-1}$ at origin and is clearly
unacceptable. This again shows that only ${\cal D}_+$ can have a solution.
In particular, the supersymmetric bound state are trustworthy even
though the quantum mechanics itself would be corrected, at small $r$,
by non-Abelian nature of such objects.

Therefore, the index problem of the above operator
is on par with that of zero mode problems around non-Abelian monopoles;
the Callias index theorem \cite{Callias:1977kg,Weinberg:1979ma,Weinberg:1979zt}
should apply. We thus anticipate that the number of zero energy bound
states is additive; when the core state is composed of many centers of
charges $\g_{c,A}$ with $ \langle\g_{c,A},\g_h\rangle\CK_0 >0$,  the number of
the bound state of the probe dyon is the naive one,
\begin{equation}
2|\langle \g_c,\g_h\rangle|=\Big|\sum_A 2\langle \g_{c,A},\g_h\rangle\Big| \ ,
\end{equation}
since $\g_c=\sum_A \g_{c,A}$.

\section{Wall-Crossing from Moduli Dynamics}

\subsection{Primitive Wall-Crossing: $\g_c+\g_h$}

So far, we ignored the precise supermultiplet structures; Our
approximation allowed us to treat the supermultiplet structure of
the core state as a separate sector, while we extracted only
partial sector of the probe dyons which would have been responsible
for building a half-hypermultiplet. More generally, the probe dyon
can come with higher spin states, such as $N=2$ vector multiplet or higher, so we may
 decompose the Hilbert space of the combined core-probe system as
\begin{equation}\label{Hilbert}
{\cal H}_\text{core}\otimes {\cal H}_\text{probe}^\text{reduced}\otimes{\cal H}_\text{moduli dynamics} \ .
\end{equation}
The reduced Hilbert space denotes part of the free Hilbert space of
a BPS particle that multiplies the half-hypermultiplet,
\begin{equation}\label{probeH}
{\cal H}= {\cal H}^\text{reduced}\otimes\left([{ 1/2}] \oplus 2 [{ 0}] \right) \ .
\end{equation}
When the probe dyon is in the half-hypermultiplet,\footnote{
Recall that usual hypermultiplet forms when the CTP conjugate
states are taken into account.}
${\cal H}^\text{reduced}_\text{probe}$ would have only one state,
while in the vector multiplet, it would be the angular momentum
1/2 Hilbert space, etc.

The decomposition (\ref{Hilbert}) can be understood easily. The
core part of the Hilbert space is inert, so can be treated as
non-dynamical. Of the probe, the half-hypermultiplet part are
generated by the universal  would-be Goldstino modes which
become no longer free due to the presence of the core state.
Instead they participate in the moduli dynamics we constructed
and thus belong to ${\cal H}_\text{moduli dynamics}$. Note that these
four modes would become free at $r=\infty$, regaining its
nature as Goldstino. The remaining part ${\cal H}^\text{reduced}_\text{probe}$
accounts for extra degeneracies and spin content of the
probe supermultiplet, which should represent additional
structure on top of the low energy dynamics.

On the other hand,  the second helicity trace (\ref{2ndh}), which is the
relevant index for $N=2$ theories,
takes value
\begin{equation}
\Omega\left([j]\otimes\left([{ 1/2}] \oplus 2 [{ 0}]\right)\right)=(-1)^{2j}(2j+1)
\end{equation}
for the irreducible angular momentum multiplet $[j]$, and can also be
expressed as
\begin{equation}
\Omega\left({\cal H}\right)= \text{tr}_{{\cal H}^\text{reduced}}(-1)^{2j_3} \ .
\end{equation}
The degrees of freedom for the core state does not participate
in the dynamics, so we have the decomposition
\begin{eqnarray}
&&\Omega\left({\cal H}_\text{core}\otimes {\cal H}_\text{probe}^\text{reduced}
\otimes{\cal H}_\text{moduli dynamics}\right)\nonumber\\
\nonumber\\
&=&\Omega\left({\cal H}_\text{core}\right)\times
\text{tr}_{{\cal H}_\text{probe}^\text{reduced}}(-1)^{2j_3}\times
\text{tr}_{{\cal H}_\text{moduli dynamics}}(-1)^{2J_3}\nonumber\\
&&\nonumber\\
&=& \Omega\left({\cal H}_\text{core}\right)\times
\Omega\left({{\cal H}_\text{probe}}\right)\times  \text{tr}_{{\cal H}_\text{moduli dynamics}}(-1)^{2j_3} \ .
\end{eqnarray}
Combining with the supersymmetric bound state we found above, this
reproduces the primitive wall-crossing formula of Denef,
\begin{equation}
\Delta \Omega(\g_c+\g_h)= -(-1)^{2|\langle \g_c,\g_h \rangle |} \;  2|\langle \g_c,\g_h \rangle |\;
\Omega(\g_c)\, \Omega(\g_h) \ .
\end{equation}

\subsection{Semi-Primitive Wall Crossing: $\g_c+n\g_h$ }

The semi-primitive wall-crossing formula of Denef and Moore
conjectures how many BPS states of charge $\g_c+n\g_h$ appears
across a MSW, for positive integer $n$
In order to compute the degeneracies of such states we must
consider $n$ number of $\g_h$ charges in the core state
background of $\g_c$. The Lagrangian would be
\begin{equation}
{\cal L}=\sum_{i=1}^n{\cal L}_{(i)} +{\cal L}_{hh} \ ,
\end{equation}
where ${\cal L}_{(i)}$ denotes the one-particle Lagrangian for
$i$-th probe dyon, all of which are of the identical form.
${\cal L}_{hh}$ captures the interaction among (identical)
probe particles.

In our approximation, the latter can be ignored as long as
the charges are such that
\begin{equation}
|\langle \g_c,\g_h\rangle| \gg |\langle \g_h,\g_{h}'\rangle| \ .
\end{equation}
In particular this is the case if the probe charges are all
mutually local, e.g., the same or proportional to each other.
Then, the latter term ${\cal L}_{hh}$
represent the second order correction to the former's first
order form and can be safely ignored. The only nontrivial
remnant  is the matter of statistics, as in any quantum mechanics
of many identical particles.

In addition, there is also a logical possibility that one-particle
BPS states of non-primitive charge $k\g_h$ exist. In
supergravity, such states are always there, since
black holes can have any quantized charges. In field theory
setting, the situation is a little unclear. In
five dimensions, multi-instanton bound state do exist
in the maximally supersymmetric Yang-Mills theory as
quantum one-particle states. However, they are tied to
the UV completion of this theory which is the mysterious
$(2,0)$ theories. In the more familiar four-dimensional
Yang-Mills setting, we are yet to see such an example.
Nevertheless, we will include the possibility that the
probe dyon of our moduli dynamics is non-primitive. Then, counting
the degeneracy of the bound states $\g_c+ n\g_h$ is basically
identical to partition of $n\g_h$ into identical halo particles
of $n\g_h=(\sum_i m_ik_i)\g_h $ with some cares on the statistics
of each dyon of charge $k_i\g_h$.
If it turns out that such non-primitive states do not exist,\footnote{The
result of the previous section is suggestive in this regard.
The bound states exist only if the Schwinger product of the two
constituent charges are nonzero. Even if we take into account
the finite core
mass, we expect that a single-particle bound state of
type $k\g+\g$ probably does not exist, which in induction
suggests absence of the state of charge $k\g$ for $k\ge 2$ altogether.
An interesting question is how this feature is modified
 in the realm of supergravity, where black holes of large non-primitive charges
 appears. } we may
simply set $\Omega(k\gamma_h)=0$ for $k\ge 2$.

The question of statistics lead us to consider the intrinsic spin of the
individual probe particle in the moduli dynamics. While the quantum
mechanics by itself won't tell us about statistics of the particle,
we can invoke the usual spin-statistics relation and
instead ask about the spin. Recall that the canonical commutators,
\begin{equation}
\{\hat \psi^m,\hat\psi^n\}=\delta^{mn} \ ,
\end{equation}
implies that the spatial rotation generators of $SU(2)_L$  acting on  the wavefunction are
\begin{equation}
-\frac{i}{4}\,[\hat\psi^a,\hat\psi^b]-\frac{i}{4}\,\epsilon_{abc}\,[\hat \psi^c,\hat\psi^4]=
\frac{1}{2}\,\e_{abc}\left(\begin{array}{cc}
 0  &0\\
0 & \sigma^a
\end{array}\right) \ .
\end{equation}
This shows that the 4-component wavefunction, $\Psi$, consists
of a single spin doublet ${\cal V}$ in the lower half and a pair of
spin singlet states combined into the upper half part, ${\cal U}$.
Recall that the bound states can appear only in the $\CV$ sector;
the supercharge $Q_4$ is effectively positively definite on $\CU$
as we saw in section 4.3. Therefore the BPS bound state of
a (half-)hypermultiplet probe and the core always involve
of a spin 1/2 wavefunction.

More generally, the probe might be in a bigger multiplet, where
${\cal H}^\text{reduced}_\text{probe}$ is also part of the data
that enters the probe dynamics although we simply factored it out.
Taking into account the latter, we can see that the probe particle  can be
seen as a particle of spin content in the moduli quantum mechanics
\begin{equation}
{\cal H}^\text{reduced}_\text{probe}\otimes([1/2]\oplus2[0]) \ ,
\end{equation}
but the BPS bound state appears only
in the sector ${\cal H}^\text{reduced}_\text{probe}\otimes[1/2]$.
For example, if ${\cal H}^\text{reduced}_\text{probe}=[S]$,
the total spin of the probe dyon that is involved in the
bound state formation is $S\pm1/2$. Therefore, {\it as far as
supersymmetric bound state formation goes, that the probe
dyon can be treated as if it is Boson or Fermion for
$2S$ odd or even, respectively.}

Such assignment of statistics is precisely what we expect on the field theory ground:
Note that $S=0$ correspond to the hypermultiplet while $S=1/2$
to the vector multiplet. When one construct BPS dyons in the weakly
coupled theory, the simplest method is to excite massive electrically
charged and $L^2$-normalizable modes around magnetic soliton \cite{Lee:1998nv}. When
the charged field is in the hypermultiplet, the relevant excitations
arise all from the Dirac field and the Fermi statistics rule when we
try to construct the dyons. For a vector multiplet, additional
modes arise both from the vector field, so the Bosonic statistics
become dominant. This naive construction
works verbatim for $N=4$ Yang-Mills theories, while for $N=2$
only slightly modified (i.e., degeneracy shift by unit) as seen from more rigorous index computation
\cite{Stern:2000ie,Denef:2002ru}. When we phrase the  $N=2$ result in terms of
vector multiplet contributions vs. hypermultiplet contributions,
we see the above statistics assignment emerging.

Interestingly, this statistics is correlated with
the sign of index $\Omega$ of the probe dyon since
\begin{equation}
\Omega\Big[[S]\otimes([{ 1/2}] \oplus 2 [{ 0}])\Big]=(-1)^{2S}(2S+1) \ .
\end{equation}
Thus, in the context of our probe moduli dynamics, probe dyons with
positive $\Omega$ should behave as Fermions, while probe dyons with
negative $\Omega$ should behave as Bosons.
More generally, ${\cal H}^\text{reduced}_\text{probe}$ can be a
direct sum of more than one spin sectors. We write
\begin{equation}
{\cal H}^\text{reduced}_\text{probe}=\oplus_\sigma [S_\sigma]={\bf R}_+\oplus {\bf R}_- \ ,
\end{equation}
with ${\bf R}_\pm$ denoting the decomposition according to the sign $(-1)^{2S_\sigma}$.
Thus,
\begin{equation}
\Omega_\text{probe}=\text{dim}{\bf R}_+ - \text{dim}{\bf R}_- \ .
\end{equation}
For the purpose of the moduli quantum mechanics here, then,
we effectively have $ \text{dim}{\bf R}_+$ Fermions
and $ \text{dim}{\bf R}_-$ Bosons of the same probe charge.

Once this statistics issue is cleared, one can construct
the generating function for the index $\O(\g_c+n\g_h)$ as follows
\begin{align}
  \sum_{n=0}^{\infty} \O(\g_c+n \g_h) q^n = \O(\g_c) \cdot \text{Tr}\Big[ \big( -1 \big)^{2J_3}
  q^N \Big] \ .
\end{align}
We used here the notation Tr to emphasize that it
is performed also over the dyons of various charges $k\g_h$ as
well as over the individual Fock space with the number operator $N$ that
counts the multiple probe dyons of the same charge. Let us split
the number operator $N=\sum_{k,j^3_\text{ext},j^3_\s} k N^B_{k,j^3_\text{ext},j^3_\s}+
\sum_{k,j^3_\text{ext},j^3_\s} k N^F_{k,j^3_\text{ext},j^3_\s}$ with $N^B$ for bosons
and $N^F$ for fermions. Here $\big|j^3_\text{ext}\big| \leq \big|\langle \g_c, k \g_h \rangle \big | -\frac12 $
and $\big| j^3_\s \big| \leq S_\s $.
The relevant trace then becomes
\begin{eqnarray}
 && \text{Tr}\Big[ \big( -1 \big)^{2J_3} q^N \Big] \nonumber\\
  &=&\sum_{N^{B/F}_{k, j^3_\text{ext},j^3_\s}} \ (-1)^{\sum_{k,j^3_\text{ext},j^3_a}
  (2j^3_\text{ext} + 2j^3_\s) \big( N^B_{k,j^3_\text{ext},j^3_\s} +
  N^F_{k,j^3_\text{ext},j^3_\s} \big)}
  q^{\sum_{k,j^3_\text{ext},j^3_\s}
  k \big( N^B_{k,j^3_\text{ext},j^3_\s} +N^F_{k,j^3_\text{ext},j^3_\s}\big)}\ ,  \nonumber
 \end{eqnarray}
which can be summed explicitly as
\begin{eqnarray}
  && \prod_{k} \prod_{j^3_\text{ext},j^3_\s} \bigg(
  \sum_{N^B=0}^\infty \Big[ (-1)^{2k|\langle \g_c, \g_h \rangle|} q^k \Big]^{N^B} \bigg)
  \cdot
  \prod_{k} \prod_{j^3_\text{ext},j^3_\s} \bigg(
  \sum_{N^F=0}^1\Big[ - (-1)^{2k|\langle \g_c, \g_h \rangle|}
  q^k \Big]^{N^F }\bigg)
  \nonumber \\
  &= & \prod_k \ \Big[ 1 - (-1)^{2k|\langle \g_c, \g_h \rangle|} q^k \Big]^{
  \text{dim}( j_\text{ext} ) \cdot \big(
  \text{dim}( {\bf R}_+) -
  \text{dim}( {\bf R}_-) \big)} \nonumber \\
  &= & \prod_k \ \Big[ 1 - (-1)^{2k\langle \g_c, \g_h \rangle} q^k \Big]^{
  2| \langle \g_c, k\g_h \rangle| \O(k\g_h) } \ .
\end{eqnarray}
It shows that the generating function is
\begin{eqnarray}\label{gen1}
   \sum_{n=0} \O(\g_c+n \g_h) q^n = \O(\g_c)\prod_{k=1} \
  \Big[ 1 - (-1)^{2k\langle \g_h, \g_c \rangle} q^k \Big]^{
  2 k | \langle \g_h, \g_c \rangle| \O(k\g_h) }  \ .
\end{eqnarray}
This is precisely the semi-crossing wall-crossing formula conjectured
by Denef and Moore \cite{Denef:2007vg}, provided that the one-particle states of charge
$\gamma_c+n\gamma_h$ are absent on the other side of the wall. Note that
the latter assumption is guaranteed by our moduli dynamics. Thus,
by staying near the walls of marginal stability and adjusting
the probe dyon to be much lighter than the core, we have derived
the semi-primitive wall-crossing formulae from the first principle.

\section{Conclusion and Discussion}

We have derived a $\CN=4$ supersymmetry low energy dynamics that govern
probe dyons interacting with relatively heavy core states, in the
long distance approximation. The proximity of the Coulomb vacuum to
the marginal stability wall acts as a crucial control parameter
that allows this non-relativistic quantum mechanical description, and
we were able to reproduce the conjectured primitive and semi-primitive
wall-crossing formulae for Seiberg-Witten theory dyons.

An important technological step here was to incorporate the potential
energy of the probe particles, due to the core state, into the
supersymmetric quantum mechanics. Because the latter comes with
different bosonic and fermionic degrees of freedom, a
nonconventional form of the supersymmetric low energy theory
emerged, but in a manner consistent with the BPS structure of the
underlying $N=2$ field theory in question.

As we mentioned early on, our approximation scheme was inspired by the
notion of framed BPS state in presence of a line operator.
See Appendix C for a short review on line operator in relation to the wall-crossing.
In a sense the line operator provides a setting
where our computation would become an exact description and can aid
evaluation of the line operator expectation values. The vacuum expectation of
line operator is in effect a $(-1)^F$ weighted trace over the Hilbert space with
a particular charge object $\Gamma$ inserted as an external object,
\begin{eqnarray}
  \langle L_\G \rangle = \text{Tr}_{\CH_{\G}}\Big[
  (-1)^F e^{-2\pi R \hat H }\Big]\ ,
  \qquad \hat H = \big\{ \CQ_\z^\dagger, \CQ_\z \big\} \ ,
\end{eqnarray}
where $\CQ_\zeta$ denote the supercharges preserved by the line operator.
It was conjectured that this observable can be expanded into
\begin{align}\label{line2}
  \langle L_\G \rangle_{\g_h} =\sum_{\g_h} \O(\G+\g_{h})  \CX_{\g_h} \ ,
\end{align}
where $\CX_{\g_h}$'s are the  Darboux coordinates of \cite{Gaiotto:2008cd}.
The semi-classical analysis on the conjectured form of $\langle L_\G \rangle$
would be interesting and illuminating as in Ref.~\cite{Chen:2010yr}.
As noted by Gaiotto et.al \cite{Gaiotto:2008cd,Gaiotto:2009hg}, this asserts the much needed
continuity property of $\CX$'s over the vacuum moduli space
that plays a central role justifying
KS formalism in the context of $N=2$ Seiberg-Witten theory. Our low
energy quantum mechanics is consistent with this claim since
\begin{eqnarray}
&& \O(\G+\g_h,\z) \CX_{\g_h}(\z,R) \nonumber\\ \nonumber\\
&=&e^{-2\pi \text{Re}[ \z^{-1}Z_{\g_h} ] }
  \text{tr}_{{\G+g_h}}
  \big[ (-1)^F e^{-2\pi R H_\text{moduli} - i \th \cdot Q } \s(Q) \big]\times (\cdots)\ ,
\end{eqnarray}
where the first two terms follows from discussions in section 2, while
$\sigma(Q)$ denotes the quadratic refinement, as argued in
Ref.~\cite{Gaiotto:2010be}. The trace is over the quantum mechanical
Hilbert space for the
charge $\G+\g_h$, while the parenthesis denotes subleading loop
contribution in the given charge sector.

An important generalization of our analysis is to study
the wall-crossing phenomena in the $N=2$ supergravity.
In fact, the formalism we developed is more natural
for the supergravity system, since the horizon provides natural
cut-off at short distance and renders the Abelian description
of the core state exact. That is, one can hide the any potential
subtlety associated with the Coulombic centers behind the horizon.
Quantum mechanical description of more than one extremally charged
black hole has been studied previously, but only in the context
of same charge black holes, which is a particular limit of our
dynamics without potential terms. We are poised to consider
many black holes with mutually non-local and interacting center,
and elevate Denef's old discussion black hole halos to fully
quantum level.

In both field theory and the supergravity version of such a low
energy quantum mechanics, there is a simpler way to
count bound states. As long as the true moduli space
defined by ${\cal K}=0$ is compact,
the relevant supercharge would be Fredholm, and one could compute the index
by concentrating on the true moduli space defined by ${\cal K}=0$.
The quantum mechanics then would reduce to a supersymmetric
Landau level problem on a curved $2n$ dimensional manifold,
and can be presumably counted by computing the volume of this
true moduli space. A similar idea has been recently used in
\cite{deBoer:2008zn,Manschot:2010qz},
but our approach provides a rigorous derivation of such a method
and thus the precise state counting. Details of
this computation will be presented elsewhere.

Finally, though we have focused on the moduli space dynamics of framed BPS
particles in $D=4$ $N=2$ supersymmetric gauge theories, our analysis can be potentially
applied to study the wall-crossing phenomena of any supersymmetric
theories in presence of higher dimensional external objects.
One potential application is a study of the wall-crossing formulae
of the four-dimensional gauge theories in presence of a surface operator,
which has been conjectured in \cite{Gaiotto:2009fs} as
a hybrid of 2D Ceccoti-Vafa WCF \cite{Cecotti:1992rm} and
4D Kontsevich-Soibelman WCF \cite{KS}. Our analysis also would be
useful to study the wall-crossing formulae of
two-dimensional $\CN=(2,2)$ massive $\mathbb{CP}^n$ models in relation to
that of four-dimensional $\CN=2$ SQCD \cite{Hanany:1997vm,Dorey:1998yh,Dorey:1999zk,Lee:2009fc}.

\vskip 1cm
\centerline{\bf\large Acknowledgement}
\vskip 5mm
We would like to thank Nick Dorey, Kazuo Hosomichi, Ki-Myeong Lee, and Andrew Neitzke
for valuable discussions.
P.Y. is supported in part by the National Research Foundation of Korea (NRF)
funded by the Ministry of Education, Science and Technology via the Center for
Quantum Spacetime (grant number 2005-0049409) and also by
Basic Science Research Program (grant number 2010-0013526).

\vskip 2cm
\newpage
\centerline{\Large \bf Appendix}

\appendix

\section{BPS Equation for the Semiclassical Core}

This appendix reviews the BPS equation, of Seiberg-Witten
low energy theory, for long-range
Abelian fields for any given core charges.
One can easily read off $N=2$ SUSY variation rules in four dimensions
from $N=1$ SUSY variation rules in six dimensions
\begin{align}\label{SUSYgaugino}
  \d \l_A = \frac12 F_{MN} \G^{MN} \e_A \ ,
\end{align}
where $\l$ and $\e$ are six-dimensional chiral spinors,
\begin{align}
  \G^{012345} \l_A = \l_A\ , \qquad
  \G^{012345} \e_A = \e_A\ .
\end{align}
Here $A=1,2$ are the R-symmetry indices. Let us decompose
the six-dimensional gamma matrices $\G^M$ as
\begin{align}
  \G^\mu = & \g^\mu \otimes {\bf 1}_2 \ , \qquad
  \G^4 =  \g_c \otimes \t^2 \ , \qquad
  \G^5 =  \g_c \otimes \t^1 \ , \qquad
  \g^\mu = \begin{pmatrix} 0 & \s^\mu \\ \bar \s^\mu & 0 \end{pmatrix}\ ,
\end{align}
where $i \g_c = \g^{0123}$. In the above representation, the gaugino
$\l_A$ can be decomposed into $\l_A = \l_{\a A} \oplus \bar \l^{\da}_A$.
As usual, $\a,\da$ denote the 4-D chiral/anti-chiral spinor indices.
One can then rewrite (\ref{SUSYgaugino}) as
\begin{align}
  \d \l_{\a A} = \frac12 F_{\mu\nu} {\s^{\mu\nu}}_\a^{\ \b} \e_{\b A} +
  i {\s^\mu}_{\a\da} \bar \e^\da_A D_\mu \phi\ ,
  \qquad \phi = A_4 +i A_5\ .
\end{align}
With $Z_c = |Z_c|\, \z$, the core state configuration should satisfy the following
relation
\begin{align}
  \Big[ \big( Q^{A} + i \z^{-1} \bar Q^A \bar \s^{0} \big)
  \vare_{A} , \l_{B} \Big] = 0
  \end{align}
or equivalently
  \begin{align}
  - i \vec \t \vare_B \cdot \big( \vec B + i \vec E - i \z^{-1} \vec \nabla \phi \big)
  - \z^{-1} \vare_B \partial_t \phi = 0 \ ,
\end{align}
that is,
\begin{align}\label{BPSconfig}
  \vec \CF - i \z^{-1} \vec \nabla \phi = 0 \ , \qquad \partial_t \phi=0\ .
\end{align}

One quick way to show that $\z$ represents the phase factor of $Z_c$
is to look at the energy for the configuration (\ref{BPSconfig}), say,
for rank one example: performing the usual trick of completing the square
with (\ref{BPSconfig}) in mind,
one obtain  %
\begin{align}
  \CE = \frac{1}{8\pi}\int d^3{\bf x} \ \text{Im} \t \Big[
  \vec B^2 + \vec E^2 + |\vec \nabla \phi |^2 \Big] =
  \text{Re} \Big[ \z^{-1} Z_c \Big] \  ,
\end{align}
with $Z_c = P \phi_D (\infty)+ Q \phi (\infty)$. This shows that
 $\z^{-1} Z_c = |Z_c|$.

%
%
%
%

\section{More on ${\CN}=4$ Quantum Mechanics}

Here we present more  on $\CN=4$ Lagrangian
with conformal $R^3$ target manifold. Here, we first derive the
massless case with curved background and then add potential terms,
which provides an alternate path to (\ref{n=4}). Then, we spend
some time on supercharge operators and quantum Hamiltonian.

\subsection{Massless and curved}

First of all, we wish to fill the gap between sections 3.1 and 3.3 with a derivation
of massless $\CN=4$ theory onto conformally flat $R^3$, which turned out to be
regarded as a special case of theories in Ref.~\cite{Maloney:1999dv}.
In next subsection, we demonstrate that how the massive Lagrangian
of section 3.3 emerges by combining the result of section 3.1
with this massless case.
Based on the educated guess and group theoretical consideration, one
possible candidate for $\CN=4$ SUSY transformation rules are
following
\begin{align}
  \d x^a = i \eta^a_{mn} \e^m \psi^n \ , \qquad
  \d \psi^m = \eta^a_{mn} \e^n {\dot x}^a + \a \e_m \eta^a_{pq} f^{-1}
  \partial_a f \psi^p \psi^q \ ,
\end{align}
where $\a$ will be determined. Here $\eta^a_{mn}$ denotes the 't Hooft
tensor with the convention $\eta^3_{12}=\eta^3_{34}=+1$.

To start, consider a standard kinetic term for flat target manifold,
\begin{align}
  \CL^{(0)} = \frac12 f {\dot x}^a {\dot x}^a + \frac i2 f \psi^m {\dot \psi}^m\ ,
\end{align}
whose variation under the $\CN=4$ SUSY transformations is
\begin{align}\label{var0}
  \d \big( \frac12 f {\dot x}^a {\dot x}^a \big) = & \frac i2 \eta^b_{mn} \partial_b f
  \e^m \psi^n {\dot x}^a {\dot x}^a + i f \eta^a_{mn} \e^m {\dot \psi}^n {\dot x}^a\ ,
  \nonumber \\
  \d \big( \frac i2 f \psi^m {\dot \psi}^m \big) = &
  -\frac12 \eta^a_{pq} \partial_a f \e^p \psi^q \psi^m {\dot \psi}^m - i f \eta^a_{mn}
  \e^m {\dot \psi}^n {\dot x}^a -\frac i2 \eta^a_{mn} \partial_b f {\dot x}^a {\dot x}^b
  \e^m \psi^n
  \nonumber \\ &
  + i \a \eta^a_{mn} \partial_a f \psi^m \psi^n \e^p {\dot \psi}^p
  + \frac i2 \a \eta^a_{mn} f^{-1} \partial_a f \partial_l f {\dot x}^l
  \psi^m \psi^n \e^p \psi^p \  .
\end{align}
\begin{enumerate}

  \item[$\bullet$] One can reorganize the velocity-square terms in (\ref{var0})
  into
  \begin{align}\label{var1}
    \frac i2 \partial_b f \e^m \psi^n  \Big[ \eta^b_{mn}  {\dot x}^a -
    & \eta^a_{mn} {\dot x}^b \Big] {\dot x}^a
    =
    \frac i2 \e_{eab} \e_{ecd} {\dot x}^a {\dot x}^c \partial_b f \eta^d_{mn}
    \e^m \psi^n
    \nonumber \\
    =   & + \frac i2 \eta^c_{pm} \eta^e_{np}  \e_{eab} {\dot x}^a {\dot x}^c  \partial_b f
    \e^m \psi^n
    \nonumber \\
    = &
    - \frac i2 \e_{eab} \cdot {\dot x}^a \partial_b f  \eta^e_{np} \d \psi^n \psi^p
    - \frac i2 \a {\dot x}^a f^{-1} \partial_a f \partial_b f
    \eta^b_{mn}\psi^m \psi^n \e^p \psi^p
    \nonumber \\ &
    + \frac i6 \a f^{-1} \partial_a f \partial_a f \e_{mnpq} \psi^m \psi^n \psi^p
    \d \psi^q \ .
  \end{align}

  \item[$\bullet$] The first term in the last equality of (\ref{var1}) implies
  that we have to add the following term
  \begin{align}\label{var2}
    \d \big( + \frac i4 \e_{abc} {\dot x}^a & \partial_b f \eta^c_{mn} \psi^m \psi^n \big)
    =  + \frac i2 \e_{abc} {\dot x}^a \partial_b f \eta^c_{mn} \d \psi^m \psi^n
    \nonumber \\ &
    - \frac 14 \e_{abc} \eta^a_{pq} \eta^c_{mn} \partial_b f \e^p {\dot \psi}^q \psi^m \psi^n
    - \frac 14 \e_{abc} {\dot x}^a \partial_b \partial_d f \eta^c_{mn}
    \eta^d_{pq} \e^p \psi^q \psi^m \psi^n \ .
  \end{align}

  \item[$\bullet$] Using the identities of 't Hooft tensor
  \begin{gather}
    \e_{abc} \eta^c_{mn}\eta^a_{pq} = \d_{mp} \eta^b_{nq} - \d_{np} \eta^b_{mq} +
    \d_{nq} \eta^b_{mp} - \d_{mq} \eta^b_{np}\ ,
    \nonumber \\
    \eta^d_{pq} \eta^c_{mn} + \eta^d_{pm} \eta^c_{nq} + \eta^d_{pn} \eta^c_{qm}
    + \eta^d_{ps} \eta^c_{rs} \e_{qmnr} =0 \ ,
  \end{gather}
  one can massage the second and third terms in (\ref{var2}) into followings:
  \begin{align}
    -\frac14 \e_{abc} \eta^a_{pq} \eta^c_{mn} \partial_b f \e^p {\dot \psi}^q \psi^m \psi^n
    = \frac12 \eta^a_{mn} \partial_a f \e^m \psi^n \cdot \psi^p {\dot \psi}^p -
    \frac12 \eta^a_{mn} \partial_a f \psi^m {\dot \psi}^n \cdot \e^p \psi^p\ ,
  \end{align}
  and
  \begin{align}
    - \frac 14 \e_{abc} {\dot x}^a \partial_b \partial_d f \eta^c_{mn}
    \eta^d_{pq} \e^p \psi^q \psi^m \psi^n  = &
    + \frac{1}{12} {\dot x}^a \partial_b\partial_d f
    \eta^d_{ps} \e_{abc} \eta^c_{rs} \e_{qmnr} \e^p \psi^q \psi^m \psi^n
    \nonumber \\
    = & + \frac{1}{12} \e_{mnpq} \partial^2f \psi^m \psi^n \psi^p \d\psi^q
    \nonumber \\ &
    - \frac{1}{12} {\dot x}^a \partial_a \partial_c f \eta^c_{pl}
    \e^p \psi^q \psi^m \psi^n \e_{qmnl}\ .
  \end{align}
  In summary, one can show that
  \begin{align}\label{var3}
    \d \big( + \frac i4 \e_{abc} {\dot x}^a \partial_b f \eta^c_{mn} \psi^m \psi^n \big)
    = & + \frac i2 \e_{abc} {\dot x}^a \partial_b f \eta^c_{mn} \d \psi^m \psi^n
    + \frac12 \eta^a_{mn} \partial_a f \e^m \psi^n \cdot \psi^p {\dot \psi}^p
    \nonumber \\ &
    + \frac14 \eta^a_{mn} \partial_a f \psi^m \psi^n \cdot \e^p {\dot \psi}^p
    + \frac{1}{12} \e_{mnpq}\partial^2 f \psi^m \psi^n \psi^p \d\psi^q \ .
  \end{align}

  \item[$\bullet$] Here one can determine, from the fourth term in second equality of
  (\ref{var0}) and third term in (\ref{var3}), the value of the coefficient $\a$ by
  \begin{align}
    \a = + \frac i4
  \end{align}

  \item[$\bullet$] Collecting all the results so far, one can have
  \begin{align}
     &\d \Big( \frac12 f {\dot x}^a {\dot x}^a + \frac i2 f \psi^m {\dot \psi}^m
    + \frac i4 \e_{abc} {\dot x}^a \partial_b f \eta^c_{mn} \psi^m \psi^n \Big)
    \nonumber \\ & = 
    \frac{1}{12} \e_{mnpq}\partial^2 f \psi^m \psi^n \psi^p \d\psi^q
    - \frac{1}{24}  f^{-1} \partial_a f \partial_a f \e_{mnpq} \psi^m \psi^n \psi^p
    \d \psi^q\ . \nonumber \\
  \end{align}
\end{enumerate}
%

At the end of the day, this gives the massless
$\CN=4$ non-linear sigma model therefore takes the following form
\begin{align}\label{Lag2}
  \CL^{(0)} = & \frac12 f {\dot x}^a {\dot x}^a  + \frac i2 f \psi^m {\dot \psi}^m
  + \frac i4 \e_{abc} {\dot x}^a \partial_b f \eta^c_{mn} \psi^m \psi^n
  \nonumber \\
  &
  - \frac{1}{48} \partial_a^2 f \e_{mnpq} \psi^m \psi^n \psi^p \psi^q
  + \frac{1}{96} f^{-1} (\partial_a f )^2 \e_{mnpq} \psi^m \psi^n \psi^p \psi^q \ ,
\end{align}
where the covariant derivative for fermions is defined as
\begin{align}
  \nabla_t \psi^m =
  {\dot \psi}^m + \frac12 \e_{abc} {\dot x}^a \partial_b \text{log}f \eta^c_{mn} \psi^n\ .
\end{align}
The above massless Lagrangian is invariant under the $\CN=4$ SUSY transformation
\begin{align}\label{SUSYvar0}
  \d x^a = i \eta^a_{mn} \e^m \psi^n \ , \qquad
  \d \psi_m = \eta^a_{mn} \e^n {\dot x}^a +\frac i4 \e_m \eta^a_{pq} f^{-1}
  \partial_a f \psi^p \psi^q \ .
\end{align}
This is the curved space version of (\ref{free}).

\subsection{Massive and curved}

Now we wish to add potential terms to this by twisting the supersymmetry transformation
rules. From discussion of section 3.2, it is clear that the right thing to do, at least
in the context of $\CN=1$ supersymmetry, is to shift the fermion transformation rule as
\begin{eqnarray}\label{SUSYvar}
  \d x^a = i \eta^a_{mn} \e^m \psi^n \ , \qquad
  \d \psi_m = \eta^a_{mn} \e^n {\dot x}^a + \e_m   \frac{1}{f}\left({\cal K}+\frac i4\eta^a_{pq}
  \partial_a f \psi^p \psi^q\right) \ ,
\end{eqnarray}
since the last piece multiplying $\e_m$ is nothing but the on-shell value of the
auxiliary field $b$. The corresponding Lagrangian from (\ref{n=4})
\begin{eqnarray}
  \CL& = & \frac12 f \Big[ {\dot x}^a {\dot x}^a  + i \psi^m \nabla_t \psi^m
  - \frac{1}{4!} \e_{mnpq}  \big\{ \nabla^2 f - (\partial_a \text{log} f)^2 \big\}
  \psi^m \psi^n \psi^p \psi^q \Big]
  \\& &
  - \CW_a {\dot x}^a + i \partial_b \CW_c \psi^b \psi^c
  + i f^{1/2} \partial_a (f^{-1/2}{\cal K}) \psi^a \l
  - \frac i4 \e_{abc} {\cal K}f^{-1}\partial_a f \psi^b \psi^c
  - \frac{1}{2f} {\cal K}^2\  \nonumber
\end{eqnarray}
is indeed consistent with the above massless one in (\ref{Lag2}).

%

To show that this Lagrangian is invariant under this transformation,
we split it into three parts,
 $\CL = \CL^{(0)} + \CL^{(1)} + \CL^{(2)}$, as
\begin{eqnarray}
  \CL^{(0)} &= & \frac12 f {\dot x}^a {\dot x}^a  + \frac i2 f \psi^m {\dot \psi}^m
  + \frac i4 \e_{abc} {\dot x}^a \partial_b f \eta^c_{mn} \psi^m \psi^n
  - \CW_a{\dot x}^a \nonumber \\ &&
  - \frac{1}{48} \partial_a^2 f \e_{mnpq} \psi^m \psi^n \psi^p \psi^q
  + \frac{1}{96} f^{-1} (\partial_a f )^2 \e_{mnpq} \psi^m \psi^n \psi^p \psi^q
  \ ,
  \nonumber \\
  \CL^{(1)}& = & \frac i2 f^{1/2} \partial_a \big( f^{1/2} K \big) \eta^a_{mn}
  \psi^m \psi^n \ ,
  \nonumber \\
  \CL^{(2)}& = &  - \frac12 f K^2\ ,
\end{eqnarray}
where we introduced $K\equiv f^{-1}{\cal K}$. ${\cal L}^{(0)}$ is already invariant
under (\ref{SUSYvar0}), so we have only $K$-dependence pieces in $\d \CL^{(0)}$, which is
\begin{eqnarray}
  \d \CL^{(0)}& = & - i f^{1/2} \partial_a ( f^{1/2} K) {\dot x}^a \e^m \psi_m -
  i \e_{abc} {\dot x}^a f^{1/2} \partial_b ( f^{1/2} K ) \eta^c_{mn} e_m \psi_n
  \nonumber \\ &&
  -\frac{1}{12} K \Big[  \partial_a^2 f - \frac12 f^{-1} (\partial_a f)^2 \Big]
  \e_{mnpq} \psi^m \psi^n \psi^p \e^q \ .
\end{eqnarray}
After some tedious computation, we find
\begin{eqnarray}
  \d \CL^{(1)}& =
  & \frac i2 \d \Big( f^{1/2} \partial_a (f^{-1/2}\cdot fK ) \Big)
  \eta^a_{mn} \psi^m \psi^n + i  f^{1/2} \partial_a (f^{1/2} K ) \d\psi^m \psi^n
  \nonumber \\ &= &
  \frac{1}{12} K \partial_a^2 f \e_{mnpq} \psi^m \psi^n \psi^p \e^q
  - \frac{1}{24} K f^{-1} ( \partial_a f )^2 \e_{mnpq} \psi^m \psi^n \psi^p \psi^q
  \nonumber \\ &&
  + i f^{1/2} \partial_a ( f^{1/2} K) {\dot x}^a \e^m \psi_m +
  i \e_{abc} {\dot x}^a f^{1/2} \partial_b ( f^{1/2} K ) \eta^c_{mn} e_m \psi_n \nonumber\\
&&+   \d (\frac12 f K^2 ) \ ,
\end{eqnarray}
%
%
%
which, combined with $\d\CL^{(0)}$, give us
\begin{eqnarray}
  \d \big( \CL^{(0)} + \CL^{(1)} \big) = \d \big( \frac12 f K^2 \big)
  = - \d \CL^{(2)}\ ,
\end{eqnarray}
%
as required.

\subsection{Supercharges and Hamiltonian}

Using the $\CN=4$ supersymmetric variation rules (\ref{susy-off}) ,
the N\"{o}ther charges of the $\CN=4$ supersymmetry therefore become
\begin{eqnarray}
  Q_m = - \eta^a_{mn} \psi^n (p_a + \CW_a)
  + \frac i4  \eta^a_{mn} f^{-1} \partial_a f \psi^n +
  \frac i4 \partial_a f \eta^a_{pq}
  \psi^{[p} \psi^q \psi^{m]} +  {\cal K} \psi^m \ .
\end{eqnarray}

For completeness, let us check whether the above supercharges give the
correct supersymmetric transformation rules for bosons and fermions.
One can read off from the Lagrangian (\ref{n=4})
the canonical quantization
\begin{eqnarray}
  \big[ x^a , p_b \big] = i \d^a_b\ , \qquad
  \big\{ \psi^m , \psi^n \big\} = f^{-1} \d^{mn}\ , \qquad
  \big[ p_a , \psi^m \big] = \frac i2 f^{-1} \partial_a f \psi^m \ .
\end{eqnarray}
One can show
\begin{eqnarray}
&& \big\{ - \eta^a_{mp} \psi^p (p_a + \CW_a) , \psi^n \big\}\nonumber\\
& = &
 - \eta^a_{mn} f^{-1} ( p_a + \CW_a )
 - \frac i2 \eta^a_{mp} f^{-1} \partial_a f \psi^p \psi^n \nonumber  \\
& = &
 - \eta^a_{mn } {\dot x}^a - \frac i4 \eta^a_{mp} f^{-1} \partial_a f \big\{
 \psi^p , \psi^n \big\}  +
 \frac i2 f^{-1} \partial_a f \eta^a_{np} \psi^{[m} \psi^{p]} \ ,\nonumber \\
 &= &
 - \eta^a_{mn } {\dot x}^a  + \frac i2 f^{-1} \partial_a f \eta^a_{np} \psi^{[m} \psi^{p]}
 - \frac i4  f^{-2}\partial_a f \eta^a_{mn} \ ,
\end{eqnarray}
where we used for the second equality the definition of momentum operator $p_a$
\begin{eqnarray}
  p_a + \CW_a = f {\dot x}^a + \frac i4 \e_{abc} \partial_b f \eta^c_{mn}\psi^m\psi^n\ .
\end{eqnarray}
One can also show that
\begin{eqnarray}
&&  \big\{ \frac i4 \partial_a f \eta^a_{pq} \psi^{[p} \psi^q \psi^{m]}  , \psi^n \big\}\nonumber\\
 & = & \d_{mn} \frac i4  f^{-1}\partial_a f \eta^a_{pq} \psi^p \psi^q +
  \frac i2 f^{-1}\partial_a f \eta^a_{np} \psi^{p} \psi^{m} + \frac i4 f^{-2} \partial_a f
  \eta^a_{mn}\ ,
\end{eqnarray}
where we used an identity of 't Hooft tensor
\begin{eqnarray}
  \e_{abc} \eta^b_{mn} \eta^c_{pq} = \eta^a_{mp} \d_{nq} - \eta^a_{np} \d_{mq}
  + \eta^a_{nq} \d_{mp} - \eta^a_{mq} \d_{np}\ .
\end{eqnarray}
It implies that
\begin{eqnarray}
  \big\{ Q_m , \psi_n \big\} = - \eta^a_{mn } {\dot x}^a
  + \d_{mn} f^{-1}\left({\cal  K} +\frac i4  f^{-1}\partial_a f \eta^a_{pq} \psi^p \psi^q\right)\ ,
\end{eqnarray}
while the action of supercharges on the bosons follows immediately,
\begin{eqnarray}
  \big[ Q_m , x^a \big] =  i \eta^a_{mn}  \psi^n \ .
\end{eqnarray}
These  are precisely the supersymmetry transformation rules in (\ref{susy-off}).

Finally, we wish to determine the quantum form of the Hamiltonian
using $$ Q_4^2=H \ .$$
Let us first write
$$Q_4=\psi^a(p+{\cal W})_a+\lambda ({\cal K}+Z) \ ,$$
where
$$Z=\frac{i}{2}\,\partial_a f \psi^a\lambda +\frac{i}{4}\,\e_{abc}\partial_a f \psi^b\psi^c \ .$$
Using $\{Q_4,\lambda\}=({\cal K}+Z)/f$ and
$\{Q_4,\psi^a\}=\dot x^a=f^{-1}\pi_a$, with the supercovariant momentum operator
$$\pi_a= (p+{\cal W})_a +\Gamma_a,\qquad\Gamma_a\equiv\frac{i}{2}\,\partial_b f \psi^{[b}\psi^{a]}-\frac{i}{2}\,\epsilon_{abc}\partial_bf\psi^c\lambda \ ,$$
we find
\begin{eqnarray}
\{Q_4,Q_4\}
&=&\{Q_4, \psi^a(p+{\cal W})_a+\lambda ({\cal K}+Z)\} \nonumber\\ \nonumber\\
&=&\frac{1}{f}\pi^a (p+{\cal W})_a +\frac{1}{f}({\cal K}+Z)^2\nonumber\\ \nonumber\\
&&-\psi^a [Q_4, (p+{\cal W})_a] -\lambda[Q_4, {\cal K}+Z] \ .
\end{eqnarray}
Let us separate out terms involving either ${\cal W}$ or ${\cal K}$ from the last two terms.
Using $d{\cal K}=*d{\cal W}$, we find
\begin{eqnarray}
\{Q_4,Q_4\} &=& \frac{1}{f}\pi^a (p+{\cal W})_a +\frac{1}{f}({\cal K}+Z)^2
-2i\partial_a {\cal K}\psi^a\lambda -i \e_{abc}\partial_a{\cal K}\psi^b\psi^c \nonumber\\ \nonumber\\
&&+\left(\psi^a[(p+{\cal W})_a,\psi^b] +\lambda[Z,\psi^b]\right)(p+{\cal W})_b \nonumber\\ \nonumber\\
&& +\left(\psi^a[(p+{\cal W})_a,\lambda] +\lambda[Z,\lambda]\right)Z \nonumber\\ \nonumber\\
&& +2\psi^a\lambda[(p+{\cal W})_a, Z] \ .
\end{eqnarray}
By explicit computation one can see that
\begin{eqnarray}
\left(\psi^a[(p+{\cal W})_a,\psi^b]
+\lambda[Z,\psi^b]\right)
&=&\frac{1}{f}\Gamma_b\nonumber\\ \nonumber\\
\left(\psi^a[(p+{\cal W})_a,\lambda] +\lambda[Z,\lambda]\right)&=&0 \ .
\end{eqnarray}
Since $ [(p+{\cal W})_a,\Gamma_a]=0$ upon the summation over $a$,
\begin{eqnarray}
\frac{1}{f}\pi^a (p+{\cal W})_a +\frac{1}{f}\,\Gamma_b(p+{\cal W})_b
&=& \frac{1}{f}\pi_a \pi_a -\frac{1}{f}\,\Gamma_a\Gamma_a \ .
\end{eqnarray}
Finally expanding $({\cal K}+Z)^2$ out, we complete the potential terms associated
with ${\cal K}$ from ${\cal K}^2+2{\cal K}Z$, but have a leftover piece $Z^2$.
So combining them all, we have
\begin{eqnarray}
\{Q_4,Q_4\} &=& \frac{1}{f}\pi^a \pi_a +\frac{1}{f}{\cal K}^2
-2if^{1/2}\partial_a(f^{-1/2} {\cal K})\psi^a\lambda -i \e_{abc}f^{1/2}\partial_a
(f^{-1/2}{\cal K})\psi^b\psi^c \nonumber\\ \nonumber\\
&&+\frac{1}{f}Z^2-\frac{1}{f}\,\Gamma_b\Gamma_b+2\psi^a\lambda[(p+{\cal W})_a, Z]
\end{eqnarray}
The last line can be organized in terms of the curvature
of the fermion bundle,
\begin{eqnarray}
[D_a, D_b]= F_{abmn}\psi^m\psi^n,\qquad D_a\equiv \partial_a+i\G_a \ ,
\end{eqnarray}
and has the explicit form,
\begin{eqnarray}\label{curvature}
%
-\frac12F_{abmn}\psi^a\psi^b\psi^m\psi^n
&=&\frac{1}{48}\,\left(2(\partial^2 f)-f^{-1}(\partial f)^2\right)\e_{mnkl}\psi^m\psi^n\psi^k\psi^l
\nonumber\\
&&-\frac14\,f^{-2}(\partial^2 f)+\frac{1}{8}f^{-3}(\partial f)^2 \ ,
\end{eqnarray}
Thus, the Hamiltonian $H=\{Q_4,Q_4\}/2$ is
\begin{eqnarray}
H 
= \frac{1}{2f}\pi_a \pi_a -\frac14 F_{abmn}\psi^a\psi^b\psi^m\psi^n+
\frac{1}{2f}{\cal K}^2-\frac{i}{2} \,\eta^{a}_{mn}f^{1/2}\partial_a(f^{-1/2}{\cal K})\psi^m\psi^n
\end{eqnarray}
Although $SU(2)_R$  is not manifest in the curvature piece,
it is actually $SU(2)_R$ invariant as can be seen from (\ref{curvature}).
This coincides with the classical Hamiltonian up to normal ordering;
the curvature pieces generate extra terms because quantum $\psi$'s
obey not the  Grassman algebra but the Clifford algebra.

Note that the kinetic term is slightly unconventional in its choice of
normal ordering. Because of this, the  inner product in the Hilbert
space of this quantum mechanics should be defined as
\begin{equation}
||\,\Psi||^2=\int dx^3 f\Psi^\dagger \Psi \ .
\end{equation}
More usual choice of kinetic term/inner product is related to our
convention by rescaling of the wavefunction by a factor of $f^{1/4}$.

\section{Review of KS  Invariant and Line Operator}

The idea of the framed BPS state originally arises in study of four-dimensional
$N=2$ supersymmetric theories in presence of an external particle of charge
$\G$, called line operator $L_\G$. The line operator can be characterized by
the phase factor $\z$ of its central charge $Z_\G$. Compactifying the theory on
a circle, it has been conjectured in \cite{Gaiotto:2010be} that the
vacuum expectation value of $L_\G$ can be expanded
in terms of the Darboux coordinates $\CX_\g$ with integer coefficients
\begin{align}\label{lvev}
  \langle L_\G \rangle = \sum_\g \O(\G+ \g) \CX_{\g}\ ,
\end{align}
which provides us a direct physical interpretation of Darboux coordinates.
Each integer coefficient $\O(\G+\g)$ here represents the supersymmetric index of
a framed BPS state of charge $\g$ bounded to $L_\G$.
The Darboux coordinates are very useful to compute the hyperK\"ahler metric on the
Coulomb branch of four-dimensional theories on a circle.

The expectation value of the line operator depends on both $\z$ and the
Coulomb branch parameter $a$ in four-dimensional theories. Due to
the fact that the physical observable $\langle L_\G \rangle$
should not have any discontinuities as $\z$ and $a$ change,
important consequences of (\ref{lvev}) are that one can understand
how the Kontsevich-Soibelman invariant naturally arises, and that
provides the origin of the thermodynamic Bethe ansatz equation the
Darboux coordinates should satisfy.

Let us now review in this section the central importance of
semi-primitive wall-crossing formula to derive the
Kontsevich-Soibelman BPS invariant in the context of line operators.
For more details, it is referred to \cite{Gaiotto:2010be}.

As discussed in the main context, the Witten index $\O(\G+\g, \z)$ can jump
once the phase of central charge for a certain probe(halo) particle of $\g_h$
is parallel to that of the external particle of $\G$ denoted by $\text{arg}(\z)$.
That is, when $\z$ moves across the so-called BPS ray
$l_{h} =\big\{ \z ~\big|~ Z_{h}/\z \in R_+\big\}$, the index could have discontinuity.
One advantage on computation of the index jump in presence of line operator
is that the wall-crossing phenomena is essentially restricted to
the semi-primitive ones.

Let us now consider the vacuum expectation value of
the line operator conjectured as in (\ref{lvev})
\begin{eqnarray}\label{line3}
  \langle L_\G \rangle = \sum_\g \O(\G+\g) \CX_{\g}\ , \nonumber
\end{eqnarray}
where $\CX_\g$ satisfy a multiplication rule below
\begin{eqnarray}
  \CX_{\g_1} \CX_{\g_2} = (-1)^{2 \langle \g_1, \g_2 \rangle} \CX_{\g_1+ \g_2}\ .
\end{eqnarray}
Let us then increase the phase parameter $\text{arg}(\z)$ so that it moves across the
BPS ray $l_{h}$.
\begin{enumerate}

\item[$\bullet$] Look at the relation (\ref{distance}).
If $\langle \g_c, \g_h \rangle >0$, we have a stable bound state
between core and halo particles before $\z$ cross the BPS ray $l_{h}$.
Then, one can reorganize (\ref{line3}) before across the ray
into the following form
\begin{eqnarray}
  \langle L_\G \rangle_- &=& \sum_{\g_c} \CX_{\g_c} \cdot \sum_{n=0}
  \O(\g_c+n\g_h) (-1)^{2 n\langle \g_c, \g_h \rangle} \CX_{\g_h}^n \ ,
  \nonumber \\
  &=& \sum_{\g_c} \O(\g_c) \CX_{\g_c}
  \prod_{n=1} \Big[ 1 - \CX_{\g_h}^n \Big]^{2 n \langle \g_c, \g_h
  \rangle \O(n\g_h) } \ .
\end{eqnarray}
Note that we used the semi-primitive wall-crossing formula (\ref{gen1})
for the last equality. Since we loose the Fock space of halo particles after across
the ray $l_{h}$, one can say that
\begin{eqnarray}
  \langle L_\G \rangle_+ = \sum_{\g_c} \O_{\g_c}  \CX_{\g_c}\ .
\end{eqnarray}
One can therefore conclude that, since $\langle L_\G \rangle$
should be continuous across the ray, $\CX_{\g_c}$ is required to jump across the wall by
the amount
\begin{eqnarray}
  \CX_{\g_c} \ \to \ \CX_{ \g_c}
  \prod_{n=1} \Big[ 1 - \CX_{\g_h}^n \Big]^{
  2 n  \langle \g_h, \g_c \rangle \O(n\g_h) } = \prod_{n=1} \CK_{n\g_h}^{\O(n\g_h)}
  (\CX_{\g_c})\ ,
\end{eqnarray}
where
\begin{align}\label{jump}
  \CK_{\g_h}(\CX_{\g_c}) = \CX_{\g_c} \Big[ 1- \CX_{\g_h} \Big]^{
  2 \langle \g_h, \g_c \rangle } \ .
\end{align}
It is noteworthy here that this is the desired discontinuity how
the Darboux coordinate $\CX_\g$ jumps across the BPS ray $l_{h}$.

\item[$\bullet$] Let us now in turn consider the
converse, i.e., $\langle \g_c, \g_h \rangle <0$. According to (\ref{distance}),
there is no stable bound state between the core and halo particle before the
$\z$ across the BPS ray $l_{h}$.
Then, one can rewrite (\ref{line3}) before across the ray as
\begin{eqnarray}
  \langle L_\G \rangle_- = \sum_{\g_c} \O_{\g_c}  \CX_{\g_c}\ .
\end{eqnarray}
Since we gain the Fock space of halo particles after across
the ray $l_{h}$, one can say that
\begin{eqnarray}
  \langle L_\G \rangle_+ &=& \sum_{\g_c} \O(\g_c) \CX_{\g_c}
  \prod_{n=1} \Big[ 1 - \CX_{\g_h}^n \Big]^{- 2n \langle \g_c, \g_h
  \rangle \O(n\g_h) } \ .
\end{eqnarray}
$\CX_{\g_c}$ is again required to jump across the wall by
the same amount
\begin{eqnarray}
  \CX_{\g_c} \ \to \ \CX_{ \g_c}
  \prod_{n=1} \Big[ 1 - \CX_{\g_h}^n \Big]^{
  2n  \langle \g_h, \g_c \rangle \O(n\g_h) } = \prod_{n=1} \CK_{n\g_h}^{\O(n\g_h)}
  (\CX_{\g_c})\ ,
\end{eqnarray}
which is the same to (\ref{jump}).

\end{enumerate}
Let us now consider two chambers of $\CM_\text{Coulomb}\times \mathbb{C}^*$,
the Coulomb branch and $\z$-plane, separated by walls of marginal stability.
The physical observable $\langle L_\G  \rangle$ should not depend on choice of
a path connecting those two chambers. The different paths however in
general cross different set of walls of marginal stability.
One can therefore conclude, from the fact that there are infinitely many
possible line operators, that a path-ordered product of transformations below
\begin{align}
  \CI= \prod_{\g_h}^\curvearrowleft \prod_n \ \CK_{n\g_h}^{\O(n\g_h)}
\end{align}
defines an invariant over the Coulomb branch $\CM_\text{Coulomb}$.
$\CI$ is indeed the so-called Kontsevich-Soibelman invariant.

\end{document}